\documentclass[camera,letterpaper,nomarginnotes,nonarrowgutter]{jpaper}
\newif\ifdraft
\draftfalse

\usepackage{tikz}
\usepackage{mathptmx} % This is Times font
\usepackage{amsmath,amssymb,amsfonts}
\usepackage{comment}
\usepackage{graphics}
\usepackage{fancyhdr}
\usepackage{booktabs}
\usepackage{pifont}
\usepackage[normalem]{ulem}
\usepackage{cite}
\usepackage{hhline}
\usepackage{xcolor}
\usepackage{setspace}
\usepackage{textcomp}
\usepackage{float}
\usepackage{enumitem}
\usepackage{afterpage}
\usepackage{graphicx}
\usepackage[htt]{hyphenat}
\usepackage{makecell}
\usepackage{xspace}
\usepackage{listings}
\usepackage{multirow}
\usepackage{balance}
\usepackage{glossaries} % acronyms
\usepackage{siunitx}
\usepackage{duckuments} %duck figures
\usepackage{dblfloatfix} %figure placement to the bottom 
\usepackage{subcaption}
\usepackage{rotating}
\usepackage[us,12hr]{datetime}
\usepackage[en-GB, useregional=numeric]{datetime2}

\usepackage[compact]{titlesec}
\usepackage[colorinlistoftodos,prependcaption,textsize=small]{todonotes} 
\usepackage{setspace}
\usepackage{dblfloatfix}    % To enable figures at the bottom of page
\usepackage{algorithm2e}
\usepackage{soul}
\usepackage{bm}

\usepackage[bookmarks=true,breaklinks=true,hidelinks]{hyperref}

\newif\ifarxiv 
\arxivtrue

\newif\ifcameraready
\camerareadyfalse

\newif\ifrev
\revfalse

\newif\ifpagenumbers
\pagenumberstrue

\newcounter{version}
\ifcameraready
\else
\fi

\title{\huge Understanding the Security Benefits and Overheads of Emerging Industry Solutions to DRAM Read Disturbance}

\newcommand{\affilETH}[0]{\textsuperscript{\S}}
\newcommand{\affilETU}[0]{\textsuperscript{$\dagger$}}
\author{
{O\u{g}uzhan Canpolat\affilETH\affilETU}\qquad
{A. Giray Ya\u{g}l{\i}kç{\i}\affilETH}\qquad
{Geraldo F. Oliveira\affilETH}\qquad
{Ataberk Olgun\affilETH}\qquad\\
{O\u{g}uz Ergin\affilETU}\qquad
{Onur Mutlu\affilETH}\qquad\\
{\affilETH \emph{ETH Z{\"u}rich}} \qquad \affilETU \emph{TOBB University of Economics and Technology}
}

\newcommand{\rfmbo}[0]{RFM+BO}
\newacronym{rfmbo}{\rfmbo{}}{refresh management with back-off support}

\newcommand{\rfmth}[0]{RFM_{th}}
\newcommand{\rfmab}[0]{RFM_{ab}}
\newcommand{\rfmsb}[0]{RFM_{sb}}
\newcommand{\nrh}[0]{N_{RH}}
\newcommand{\aboth}[0]{N_{BO}}
\newcommand{\taboact}[0]{t_{ABO_{ACT}}}
\newcommand{\tbodelay}[0]{t_{BackOffDelay}}
\newcommand{\bonrefs}[0]{N_{Ref}}
\newcommand{\bonacts}[0]{N_{Delay}}

\newcommand{\maxact}[0]{\text{MAX}_{\text{ACT}}}
\newcommand{\maxrfm}[0]{\text{MAX}_{\text{RFM}}}
\newcommand{\dallref}[0]{D_{\text{refresh}}}
\newcommand{\trfmperiod}[0]{T_{\text{RFM}}}

\newcommand{\trcd}[0]{t_{RCD}}
\newcommand{\tras}[0]{t_{RAS}}
\newcommand{\trp}[0]{t_{RP}}
\newcommand{\trc}[0]{t_{RC}}
\newcommand{\trefi}[0]{t_{REFI}}
\newcommand{\trefw}[0]{t_{REFW}}
\newcommand{\trfc}[0]{t_{RFC}}
\newcommand{\trrd}[0]{t_{RRD}}
\newcommand{\trfm}[0]{t_{RFM}}
\newcommand{\trtp}[0]{t_{RTP}}
\newcommand{\twr}[0]{t_{WR}}

\newacronym{maxact}{$\maxact{}$}{maximum activation commands possible within a refresh window}
\newacronym{maxrfm}{$\maxrfm{}$}{maximum refresh management commands possible within a refresh window}
\newacronym{nrh}{$\nrh{}$}{\emph{minimum hammer count to induce the first bitflip}}
\newacronym{dallref}{$\dallref{}$}{time taken to complete refresh commands within a refresh window}
\newacronym{trfmperiod}{$\trfmperiod{}$}{the minimum time needed between two consecutive \gls{RFM} commands targeting the same bank}

\newacronym{trefw}{$\trefw{}$}{refresh window}
\newacronym{trefi}{$\trefi{}$}{refresh interval}
\newacronym{trfm}{$\trfm{}$}{refresh management latency}
\newacronym{trfc}{$\trfc{}$}{refresh latency}

\newacronym{rfm}{RFM}{refresh management}
\newacronym{prfm}{PRFM}{Periodic RFM}
\newacronym{rfmth}{$\rfmth{}$}{\emph{bank activation threshold to issue an RFM command}}
\newacronym{rfmab}{$\rfmab{}$}{all bank refresh management}
\newacronym{rfmsb}{$\rfmsb{}$}{same bank refresh management}
\newacronym{prac}{PRAC}{Per Row Activation Counting}
\newacronym{abo}{$ABO$}{Alert Back-Off}
\newacronym{aboth}{$\aboth{}$}{the back-off threshold}
\newacronym{taboact}{$\taboact{}$}{\emph{the window of normal traffic}}
\newacronym{tbodelay}{$\tbodelay{}$}{the delay until a new back-off can be initiated}
\newacronym{trc}{$\trc{}$}{row cycle time}

\newacronym{trcd}{$\trcd{}$}{row activation latency}
\newacronym{tras}{$\tras{}$}{charge restoration latency}
\newacronym{trp}{$\trp{}$}{precharge latency}
\newacronym{trrd}{$\trrd{}$}{the minimum time window between two row activation commands targeting different banks}

\newcommand{\tfilter}[0]{T_{Filter}}
\newacronym{tfilter}{$\tfilter{}$}{the filtering threshold}

\newcommand{\act}[0]{ACT}
\newcommand{\pre}[0]{PRE}
\newcommand{\refresh}[0]{REF}
\newcommand{\wri}[0]{\texttt{WR}}
\newcommand{\rd}[0]{\texttt{RD}}

\newcommand{\apa}[0]{\texttt{APA}}

\newcommand{\pum}[0]{\texttt{PuM}}

\usepackage[shortcuts]{extdash}

\newacronym{iqr}{$IQR$}{inter-quartile range}
\newacronym{act}{\act{}}{activate}
\newacronym{pre}{\pre{}}{precharge}
\newacronym{ref}{\refresh{}}{refresh}
\newacronym{wr}{\wri{}}{write}
\newacronym{rd}{\rd{}}{read}
\newacronym{pum}{\pum{}}{Processing-using-Memory}
\newacronym{apa}{\apa{}}{\act{} $\rightarrow$ \pre{} $\rightarrow$ \act{}}
\newacronym{jedec}{JEDEC}{Joint Electron Device Engineering Council}
\newacronym{tcl}{$t_{CL}$}{column access latency}
\newacronym{tcwl}{$t_{CWL}$}{column write latency}

\newacronym{puf}{PUF}{physical unclonable function}
\newacronym{trn}{TRN}{true random number}
\newacronym{taggon}{$t_{AggOn}$}{the time that an aggressor row stays active, {i.e., aggressor row's on-time}}

\definecolor{gfored}{rgb}{0.580, 0.050, 0.211}
\definecolor{ao}{rgb}{0.007, 0.520, 0.867}
\definecolor{moegi}{rgb}{0.357, 0.537, 0.188}
\definecolor{jl}{rgb}{1.0, 0.2, 0.8}
\definecolor{brown(web)}{rgb}{0.65, 0.16, 0.16}
\definecolor{bisque}{rgb}{1.0, 0.89, 0.77}
\definecolor{nbs}{rgb}{0.88, 0.07, 0.37}
\definecolor{yt}{rgb}{0.58, 0.44, 0.86}
\definecolor{iy}{rgb}{0.0, 0.36, 0.05}
\definecolor{burntorange}{rgb}{0.8, 0.33, 0.0}
\newcommand{\ignore}[1]{}

\ifdraft
    
    \newcommand{\param}[1]{\textcolor{red}{#1}} 

\else
\fi

\definecolor{frenchblue}{rgb}{0.19, 0.55, 0.91}

\usepackage[most]{tcolorbox} 
\tcbset{before skip=1.5pt, after skip=4pt}

\newtcolorbox[auto counter]{obsx}[3][]{%
    colframe = #2!45,
    colback  = #2!10,
    coltitle = #2!20!black, 
    colbacktitle=#2!20,
    coltitle=black,
    fonttitle=\bfseries, 
    title=#3~\thetcbcounter.\ ,
    enhanced,
    attach boxed title to top left={yshift=-2.8mm, xshift=0.15cm},
    bottom=-2.2pt,
    #1% 
}

\usepackage[most]{tcolorbox} 
\newtcolorbox[auto counter]{tkx}[2][]{%
    enhanced, breakable, center title,
    colframe = #2!45,
    colback  = #2!10,
    colbacktitle=#2!20,
    left=-0.5pt,
    right=-0.5pt,
    bottom=-2pt,
    top=-0.25pt,
    #1% 
}
\newcounter{obs}
\definecolor{amber}{rgb}{1.0, 0.49, 0.0}
\definecolor{awesome}{rgb}{1.0, 0.13, 0.32}
\definecolor{dollarbill}{rgb}{0.52,0.73,0.4}
\definecolor{moegi}{rgb}{0.357, 0.537, 0.188}
\definecolor{burgundy}{rgb}{0.5, 0.0, 0.13}
\definecolor{ballblue}{rgb}{0.13, 0.67, 0.8}
\definecolor{ups-truck}{rgb}{0.53, 0.28, 0.21}
\definecolor{airforceblue}{rgb}{0.36, 0.54, 0.66}
\definecolor{cadmiumgreen}{rgb}{0.0, 0.42, 0.24}
\definecolor{darkcyan}{rgb}{0.0, 0.55, 0.55}
\definecolor{caribbeangreen}{rgb}{0.0, 0.8, 0.6}
\definecolor{flamingopink}{rgb}{0.99, 0.56, 0.67}
\definecolor{jazzberryjam}{rgb}{0.65, 0.04, 0.37}
\definecolor{mediumpersianblue}{rgb}{0.0, 0.4, 0.65}
\definecolor{coolblack}{rgb}{0.0, 0.18, 0.39}
\definecolor{bleudefrance}{rgb}{0.19, 0.55, 0.91}
\definecolor{ao}{rgb}{0.0, 0.0, 1.0}
\definecolor{babyblueeyes}{rgb}{0.63, 0.79, 0.95}
\definecolor{darkwarmgray}{rgb}{0.2,0,0}
\definecolor{brightpink}{rgb}{1.0, 0.0, 0.5}
\definecolor{iy}{rgb}{0.0, 0.36, 0.05}

\newcommand{\squishlist}{
 \begin{list}{$\circ$}
  { \setlength{\itemsep}{0pt}
     \setlength{\parsep}{0pt}
     \setlength{\topsep}{0pt}
     \setlength{\partopsep}{0pt}
     \setlength{\leftmargin}{1em}
     \setlength{\labelwidth}{1em}
     \setlength{\labelsep}{0.5em} } }

\newcommand{\squishsublist}{
\begin{list}{$\rightarrow$}
 { \setlength{\itemsep}{0pt}
    \setlength{\parsep}{0pt}
    \setlength{\topsep}{-10em}
    \setlength{\partopsep}{-3pt}
    \setlength{\leftmargin}{1em}
    \setlength{\labelwidth}{1em}
    \setlength{\labelsep}{0.5em} } }

\newcommand{\squishend}{
  \end{list}  }

\newcommand{\head}[1]{\noindent\textbf{#1.}}

\newcounter{take}
\ifdraft
\paperwidth=\dimexpr \paperwidth + 4cm\relax
\oddsidemargin=\dimexpr\oddsidemargin + 2cm\relax
\evensidemargin=\dimexpr\evensidemargin + 2cm\relax
\marginparwidth=\dimexpr \marginparwidth + 2cm\relax
\fi
\newcommand{\gf}[2]{\ifnum#1=\value{version}\textcolor{red}{#2}\else{#2}\fi}
\newcommand{\agy}[2]{\ifnum#1=\value{version}\textcolor{blue}{#2}\else{#2}\fi}
\newcommand{\atb}[2]{\ifnum#1=\value{version}\textcolor{orange}{#2}\else{#2}\fi}
\newcommand{\yct}[2]{\ifnum#1=\value{version}\textcolor{yt}{#2}\else{#2}\fi}
\newcommand{\ous}[2]{\ifnum#1=\value{version}\textcolor{purple}{#2}\else{#2}\fi}
\newcommand{\iey}[2]{\ifnum#1=\value{version}\textcolor{iy}{#2}\else{#2}\fi}
\newcommand{\om}[2]{\ifnum#1=\value{version}\textcolor{gfored}{#2}\else#2\fi}

\newcommand{\agytodo}[2]{\ifnum#1=\value{version}\todo[size=\scriptsize, linecolor=orange, bordercolor=orange, backgroundcolor=white]{\textcolor{blue}{TODO:~#2}}\else{}\fi}
\newcommand{\ycttodo}[2]{\ifnum#1=\value{version}\todo[size=\scriptsize, linecolor=orange, bordercolor=orange, backgroundcolor=white]{\textcolor{yt}{TODO:~#2}}\else{}\fi}
\newcommand{\ieytodo}[2]{\ifnum#1=\value{version}\todo[size=\scriptsize, linecolor=orange, bordercolor=orange, backgroundcolor=white]{\textcolor{iey}{TODO:~#2}}\else{}\fi}

\newcommand{\agycomment}[2]{\ifnum#1=\value{version}\todo[size=\scriptsize, linecolor=orange, bordercolor=orange, backgroundcolor=white]{\textcolor{blue}{Giray:~#2}}\else{}\fi}
\newcommand{\atbcomment}[2]{\ifnum#1=\value{version}\todo[size=\scriptsize, linecolor=orange, bordercolor=orange, backgroundcolor=white]{\textcolor{orange}{Atb:~#2}}\else{}\fi}
\newcommand{\yctcomment}[2]{\ifnum#1=\value{version}\todo[size=\scriptsize, linecolor=orange, bordercolor=orange, backgroundcolor=white]{\textcolor{yt}{Yahya:~#2}}\else{}\fi}
\newcommand{\ouscomment}[2]{\ifnum#1=\value{version}\todo[size=\scriptsize, linecolor=orange, bordercolor=orange, backgroundcolor=white]{\textcolor{purple}{Oguzhan:~#2}}\else{}\fi}
\newcommand{\gfcomment}[2]{\ifnum#1=\value{version}\todo[size=\scriptsize, linecolor=orange, bordercolor=orange, backgroundcolor=white]{\textcolor{purple}{Oguzhan:~#2}}\else{}\fi}
\newcommand{\ieycomment}[2]{\ifnum#1=\value{version}\todo[size=\scriptsize, linecolor=orange, bordercolor=orange, backgroundcolor=white]{\textcolor{iy}{Ismail:~#2}}\else{}\fi}
\newcommand{\omcomment}[2]{\ifnum#1=\value{version}\todo[size=\scriptsize, linecolor=orange, bordercolor=orange, backgroundcolor=white]{\textcolor{gfored}{Onur:~#2}}\else{}\fi}
\newcommand{\versionedparam}[2]{\ifnum#1=\value{version}{#2}\else{#2}\fi}
\providecommand{\param}[1]{\versionedparam{\value{version}}{#1}}

\newcommand{\secref}[1]{§\ref{#1}}

\newcommand{\tabref}[1]{Table~\ref{#1}}

\newcommand{\figref}[1]{Fig.~\ref{#1}}
\newcommand{\figsref}[1]{Figs.~\ref{#1}}

\newcommand{\revtag}[1]{}
\newcommand{\copied}[2]{#2}
\newcommand{\rhmemisolationrefs}[0]{\cite{fournaris2017exploiting, poddebniak2018attacking, tatar2018throwhammer, carre2018openssl, barenghi2018software, zhang2018triggering, bhattacharya2018advanced, google-project-zero, kim2014flipping, rowhammergithub, seaborn2015exploiting, van2016drammer, gruss2016rowhammer, razavi2016flip, pessl2016drama, xiao2016one, bosman2016dedup, bhattacharya2016curious, burleson2016invited, qiao2016new, brasser2017can, jang2017sgx, aga2017good, mutlu2017rowhammer, tatar2018defeating, gruss2018another, lipp2018nethammer, van2018guardion, frigo2018grand, cojocar2019eccploit,  ji2019pinpoint, mutlu2019rowhammer, hong2019terminal, kwong2020rambleed, frigo2020trrespass, cojocar2020rowhammer, weissman2020jackhammer, zhang2020pthammer, yao2020deephammer, deridder2021smash, hassan2021utrr, jattke2022blacksmith, tol2022toward, kogler2022half, orosa2022spyhammer, zhang2022implicit, liu2022generating, cohen2022hammerscope, zheng2022trojvit, fahr2022frodo, tobah2022spechammer, rakin2022deepsteal, park2016statistical, park2016experiments,lim2017active, ryu2017overcoming, yun2018study, yang2019trap, walker2021ondramrowhammer, kim2020revisiting, orosa2021deeper, yaglikci2022understanding, khan2018analysis, agarwal2018rowhammer, li2014write, ni2018write, genssler2022reliability, mutlu2023fundamentally}}

\newcommand{\mcBasedRowHammerMitigations}[0]{\cite{AppleRefInc, rh-hp,LenovoRefInc,greenfield2012throttling, kim2014flipping, kim2014architectural, aichinger2015ddr, aweke2016anvil, bains-merged, bains2015row, bains2016distributed, bains2016row, son2017making, irazoqui2016mascat, ryu2017overcoming, yang2017scanning, you2019mrloc, lee2019twice, park2020graphene, yaglikci2021blockhammer, frigo2020trrespass, kang2020cattwo, hassan2021utrr, qureshi2022hydra, saileshwar2022randomized, brasser2017can, konoth2018zebram, van2018guardion, vig2018rapid, gautam2019row, kim2022mithril, lee2021cryoguard, zhang2022softtrr, joardar2022learning, juffinger2023csi, yaglikci2022hira, saxena2022aqua, enomoto2022efficient, manzhosov2022revisiting, ajorpaz2022evax, joardar2022machine, hassan2022acase, zhang2020leveraging, loughlin2021stop, devaux2021method, han2021surround, fakhrzadehgan2022safeguard, saroiu2022price, saroiu2022configure, loughlin2022moesiprime, zhou2022ltpim, mutlu2023fundamentally, didio2023copyonflip, sharma2022areview, woo2023scalable, park2022rowhammer, wi2023shadow, kim2023ddr5, gude2023defending, guha2022criticality, france2022modeling, france2022reducing, arikan2022processor, tomita2022extracting, saxena2023pt, zhou2023dnndefender}}

\newcommand{\inDRAMRowHammerMitigations}[0]{\cite{jedec2020jesd794c, jedec2020jesd795, gomez2016dummy, seyedzadeh2017counterbased, seyedzadeh2018mitigating, yang2016suppression, yaglikci2021security, devaux2021method, marazzi2022protrr, hassan2019crow, hong2023dsac, marazzi2023rega, bennett2021panopticon, wi2023shadow}}

\newcommand{\refreshBasedRowHammerDefenseCitations}[0]{\cite{lee2019twice, seyedzadeh2017counterbased, seyedzadeh2018mitigating, kang2020cattwo, park2020graphene, kim2022mithril, kim2014architectural, bains2015row, bains2016distributed, bains2016row, aweke2016anvil, AppleRefInc, kim2014flipping, son2017making, you2019mrloc, yaglikci2021security, frigo2020trrespass, hassan2021utrr, qureshi2022hydra, devaux2021method, lee2021cryoguard, marazzi2022protrr, zhang2022softtrr, joardar2022learning}}

\makeatletter
\def\bstctlcite{\@ifnextchar[{\@bstctlcite}{\@bstctlcite[@auxout]}}
\def\@bstctlcite[#1]#2{\@bsphack
 \@for\@citeb:=#2\do{%
   \edef\@citeb{\expandafter\@firstofone\@citeb}%
   \if@filesw\immediate\write\csname #1\endcsname{\string\citation{\@citeb}}\fi}%
 \@esphack}
\makeatother

\newcounter{observation}
\newcounter{corollary}
\newcounter{takeaway}
\newcounter{claim}
\newcounter{proof}

\ifarxiv
    \pagenumberstrue
\else
    \ifcameraready
    \else 
        \newcounter{passversion}
        \pagenumberstrue
        \fancypagestyle{firstpage}
        {
            \fancyhead{}
            \fancyhead[C]{\textcolor{red}{CONFIDENTIAL DRAFT -- DO NOT DISTRIBUTE -- TO APPEAR IN DRAMSec'24} \\ \textcolor{MidnightBlue}{\emph{Version \thepassversion{}.4~---~\today, \ampmtime}} }
        }
    \fi
\fi

\makeatletter
\def\bstctlcite{\@ifnextchar[{\@bstctlcite}{\@bstctlcite[@auxout]}}
\def\@bstctlcite[#1]#2{\@bsphack
  \@for\@citeb:=#2\do{%
    \edef\@citeb{\expandafter\@firstofone\@citeb}%
    \if@filesw\immediate\write\csname #1\endcsname{\string\citation{\@citeb}}\fi}%
  \@esphack}
\makeatother 

\begin{document}
\bstctlcite{IEEEexample:BSTcontrol}

\maketitle

\newcommand{\hpcaheight}{0mm}
\ifdefined\eaopen
\renewcommand{\hpcaheight}{12mm}
\fi
\ifcameraready
    \thispagestyle{empty}
    \pagestyle{empty}
\else 
    \pagestyle{plain}
\fi

\ifcameraready
    \setcounter{version}{99}
\else
    \setcounter{version}{99}
\fi
\begin{abstract}
    \ous{6}{We} present the first rigorous security, performance, energy, and cost analyses of
    \agy{2}{the state-of-the-art on-DRAM-die read disturbance mitigation method, widely known as \gls{prac}\gf{2}{,} with respect to its description in the updated \ous{6}{(as of \agy{6}{April} 2024)} \gf{2}{JEDEC} DDR5 specification.}
    \agy{2}{Unlike prior state-of-the-art \gf{2}{that} \ous{6}{advises} the memory controller \gf{2}{to} periodically issue a DRAM command called \gls{rfm}\gf{2}{, which} provides the DRAM chip with time to perform its \ous{6}{countermeasures},
    % widely known as \gls{rfm}, 
    \gls{prac} introduces a new back-off signal. 
    \agy{4}{\gls{prac}'s back-off signal propagates from the DRAM chip to the memory controller and forces the memory controller to 1)~stop serving requests and 2)~issue RFM commands. As a result, \gls{rfm} commands are issued only when needed as opposed to periodically,}
    % This back-off signal allows the DRAM chip to force the memory controller to issue an \gls{rfm} only when needed, 
    reducing the performance overhead of \gls{rfm}.}
    % suggests that 1)~the DRAM chip accurately tracks row activation counts and asserts a back-off signal when a preventive refresh is needed, and 2)~upon receiving the back-off signal, the memory controller provides the DRAM chip with a time window to perform necessary preventive refreshes using a \gls{rfm} command. Compared to the prior specifications, \gls{prac}'s novelty is to introduce the back-off signal from the DRAM chip to the memory controller. Therefore, we refer to this mechanism as \gls{rfmbo}.}
    % \agy{1}{two new features of DRAM standards that are useful for practical and low-overhead read disturbance mitigation:} \gls{rfm} and \gls{prac}, 
    % \agy{1}{which provides the DRAM chip with the necessary time window to perform preventive refresh operations when necessary.}
    % we first provide a brief explanation of \gls{rfm} and \gls{prac}. 
    % in a producer--consumer scheme, where the system is considered to be secure as long as all preventive refresh requests that the in-DRAM solution generates (produces) are performed (consumed) before an aggressor row's activation count exceeds a threshold. 
    \agy{2}{We analyze \gls{prac} in \agy{6}{four} steps.
    First, we define \agy{7}{a security-oriented} adversarial access pattern that represents the worst-case for the security of \gls{prac}. 
    Second, we investigate \gls{prac}'s different configurations and their security implications.
    Our security analyses show that \gls{prac} can be configured for \ous{6}{secure} operation as long as no bitflip occurs before accessing a memory location \param{\ous{12}{20}} times.
    Third, we evaluate the performance impact of \gls{prac} and compare it against prior works using an open-source cycle-level simulator\ous{6}{, Ramulator 2.0}.
    Our performance analysis shows that while \gls{prac} incurs less than \ous{12}{\param{13}}\% performance overhead on benign applications for \ous{6}{today's} DRAM chips, its performance overheads can reach up to \ous{12}{\param{94}}\% (\ous{12}{\param{85}}\% on average across \param{60} workloads) for future DRAM chips \ous{6}{that are more vulnerable to read disturbance bitflips}.
    \agy{6}{Fourth, we define \agy{7}{an availability-oriented} adversarial access \om{8}{pattern that} \om{7}{exacerbates} \agy{7}{\gls{prac}'s performance overhead}
    % the performance of \gls{prac} 
    to perform a memory performance attack, \om{7}{demonstrating} that such \om{7}{an} adversarial pattern can \om{7}{hog} up to \ous{12}{\param{94\%}} of DRAM \om{7}{throughput} \om{8}{and degrade system throughput by up to \ous{12}{\param{95\%}} (\ous{12}{\param{87\%}} on average)}.} 
    We discuss \agy{7}{\gls{prac}'s} implications on future systems and foreshadow future research directions.
    \om{7}{To aid future research, we open-source our implementations and scripts at \url{https://github.com/CMU-SAFARI/ramulator2}.\agycomment{7}{Dropped ``for now''}}}
\end{abstract}

\glsresetall{}
\section{Introduction}
\label{sec:intro}

%Copied from Svard
{To ensure system \om{2}{robustness (including} reliability, security, and safety\om{2}{)}, it is critical to {maintain} {memory isolation: accessing a memory address should \emph{not} cause unintended side-effects on data stored on other addresses~\cite{kim2014flipping}. {Unfortunately}, with aggressive technology scaling, DRAM~\cite{dennard1968fieldeffect}, the prevalent {main} memory technology}, suffers from increased \emph{read disturbance}: accessing (reading) a row of DRAM cells (\om{2}{i.e., a} DRAM row) degrades the data integrity of other physically close but \emph{unaccessed} DRAM rows.}
% \emph{RowHammer}~\cite{kim2014flipping} and \emph{RowPress}~\cite{luo2023rowpress} are two prime \ous{6}{examples of} {DRAM read {disturbance}, {where} a DRAM row (i.e., victim row) can experience bitflips when a nearby DRAM row (i.e., aggressor row) is} 1)~repeatedly opened (i.e., hammered)~\rhmemisolationrefs{} or 
% 2)~kept open for a long period (i.e., pressed)~\cite{luo2023rowpress}, respectively.
\agy{6}{RowHammer~\cite{kim2014flipping} is a prime example of DRAM read disturbance, where a DRAM row (i.e., victim row) can experience bitflips when \ous{7}{at least one} nearby DRAM row (i.e., aggressor row) is repeatedly \om{7}{activated} (i.e., hammered)~\rhmemisolationrefs{} more times than a threshold, called the \om{7}{\gls{nrh}}. \emph{RowPress}~\cite{luo2023rowpress} is another prime example of DRAM read disturbance that amplifies the effect of RowHammer and consequently reduces \gls{nrh}.}

% \vspace{-0.5em}
A simple way of mitigating DRAM read disturbance is to preventively refresh potential victim rows before bitflips occur. \om{7}{Doing so} comes at the cost of potential performance degradation~\refreshBasedRowHammerDefenseCitations{}.
% However, such solutions can reduce system performance As technology node scaling exacerbates DRAM read disturbance, these mechanisms preventively refresh potential victim rows more frequently, incurring higher performance overheads. Therefore, to prevent read disturbance bitflips at low performance overheads, it is important to accurately measure each row's activation count and perform preventive refresh operations \emph{only} when necessary.
% To accurately measure row activation counts at low cost, several works~\inDRAMRowHammerMitigations{} propose maintaining a row activation counter for each row and maintain those counters in DRAM chip, leveraging DRAM's high density and low cost-per-bit. Unfortunately, DRAM chip does \emph{not} have the autonomy to allocate an arbitrary time window to perform such refresh operations as the memory controller (in processor) has fine-grained control over DRAM operations and timings. 
To provide DRAM chips with the necessary flexibility to perform preventive refreshes \ous{6}{in a timely manner}, \ous{6}{recent} DRAM standards (e.g., DDR5~\cite{jedec2020jesd795, jedec2024jesd795c}) introduce 1)~a command called \agy{7}{\emph{\gls{rfm}}}\ous{6}{\cite{jedec2020jesd795}} and 2)~a mechanism called \agy{7}{\emph{\gls{prac}}}\ous{6}{\cite{jedec2024jesd795c}}.
\gls{rfm} is a DRAM command that provides the DRAM chip with a time window (e.g., \param{\SI{195}{\nano\second}}~\cite{jedec2024jesd795c}) to perform preventive refreshes.
\ous{6}{Specifications before 2024} (e.g., DDR5~\cite{jedec2020jesd795}) \ous{6}{advise} the memory controller \ous{6}{to issue} \gls{rfm} when the number of row activations in a bank or a logical memory region exceeds a threshold\ouscomment{7}{removed ``value'' to kill an orphan} (e.g., \param{32}~\cite{jedec2024jesd795c}).
% threshold value
\agy{2}{A recent update \ous{6}{as of \agy{6}{April} 2024} \om{7}{of} \gf{2}{the JEDEC} DDR5 specification~\cite{saroiu2024ddr5, jedec2024jesd795c} introduces a new on-DRAM-die read disturbance mitigation mechanism called \gls{prac}\ouscomment{7}{added long-form earlier}.}
\ous{6}{\gls{prac} has two key features.
First, \gls{prac} maintains an activation counter per DRAM row~\cite{kim2014flipping} to accurately identify when a preventive refresh is needed.
\agy{6}{\gls{prac} increments a DRAM row's activation counter while the row is being closed, \om{7}{which increases} the latency of closing a row\agy{7}{,} \om{7}{i.e., the \gls{trp} and \gls{trc} timing parameters}.}
% These counters are updated each time a DRAM row is accessed, increasing the necessary time to access DRAM.
Second, \gls{prac}} proposes a new \emph{back-off} signal to convey \ous{6}{the need for preventive refreshes} from the DRAM chip to the memory controller, similar to what prior works propose~\cite{bennett2021panopticon, devaux2021method, yaglikci2021security, hassan2022acase, kim2022mithril}.
The DRAM chip asserts this back-off signal when a DRAM row's activation count reaches a critical value.
%1)~maintains an activation counter per DRAM row\ous{6}{\cite{kim2014flipping}} to accurately identify when a preventive refresh is needed \ous{6}{(at the cost of increased DRAM timing parameters to update these counters)} and
Within a predefined time window (e.g., \param{\SI{180}{\nano\second}}~\cite{jedec2024jesd795c}) after receiving the back-off signal, the memory controller \ous{6}{has} to issue an \gls{rfm} command so that the DRAM chip can perform the \ous{6}{necessary} preventive refresh operations.
\gls{prac} \agy{2}{aims to}
1)~\ous{6}{avoid} read disturbance bitflips \ous{6}{by} performing necessary preventive refresh\ous{4}{es} \ous{6}{in a timely manner} and
2)~\ous{6}{minimize} unnecessary preventive \ous{4}{refreshes} \ous{6}{by} accurately \ous{4}{tracking} \om{7}{each row's activation count}.
Unfortunately, \emph{no} prior work rigorously investigates the \agy{2}{impact of \gls{prac} on security, performance, energy, \ous{4}{and cost} for}
% security limitations and performance overheads of \gls{rfm} and \gls{prac} in 
modern and future systems. 

% copied from the abstract. This needs elaboration based on the key findings when they are ready.
\ous{6}{This paper performs} the first rigorous \ous{6}{analysis} of \gls{prac} \agy{2}{in \agy{6}{four} steps}.
% security, \agy{2}{performance, energy, and cost} analyses
% To do so, we first methodically define \gls{rfm} and \gls{prac}. 
% in a producer--consumer scheme, where the system is considered to be secure as long as all preventive refresh requests that the in-DRAM solution generates (produces) are performed (consumed) before an aggressor row's activation count exceeds a threshold. 
\agy{2}{First}, we define a \om{8}{security-oriented} adversarial access pattern that achieves the highest possible activation count in systems protected by \gls{prac}. 
\agy{2}{Second, we conduct a security analysis by evaluating the highest possible activation count that a DRAM row can reach under different configurations of \gls{prac}.}
Our analysis shows that \gls{prac} can be configured for \ous{6}{secure} operation \ous{6}{against \agy{6}{an \gls{nrh} value of \ous{12}{20} or higher.}}
\agy{2}{Third}, we evaluate \agy{2}{the impact of \gls{prac} on performance and energy} 
% impact of \gls{rfm} and \gls{prac}, 
using \ous{6}{Ramulator 2.0~\cite{luo2023ramulator2,ramulator2github}, \ous{6}{an open-source} cycle-level simulator extended with DRAMPower~\cite{drampower}}.
Our \agy{2}{results across \param{60} different \ous{4}{four}-core multiprogrammed benign workload mixes} show that \gls{prac} incurs \ous{6}{an average (maximum) \ous{12}{\param{9.9\%}} (\ous{12}{\param{13.1\%}}) system} performance and \ous{12}{\param{18.5\%} (\ous{12}{\param{22.7\%}}) DRAM} energy overheads for modern DRAM chips \agy{6}{with \agy{7}{any of the} \gls{nrh} values of 10K~\cite{kim2020revisiting}, 4.8K~\cite{kim2020revisiting}, and} 1K~\cite{luo2023rowpress}.
\agy{6}{These overheads are \om{8}{similar} \agy{7}{across the specified \gls{nrh} values because they are \om{8}{mainly a result} of} the increased \agy{7}{critical DRAM access latencies} \ous{7}{(i.e., \gls{trp} and \gls{trc})}.}
\ous{7}{\gls{prac}'s average (maximum) overheads reach \ous{12}{\param{10.1\%} (\param{13.3\%}), \param{11.8\%} (\param{15.7\%}), and \param{84.7\%} (\param{94.0\%})} for performance and \ous{12}{\param{18.8\%} (\param{22.9\%}), \param{20.8\%} (\param{25.7\%}), and \param{13x} (\param{18x})} for DRAM energy on future DRAM chips with \gls{nrh} values of 128, 64, and \ous{12}{20}, respectively.
We attribute these large overheads to many preventive refresh operations being performed (even under benign workloads) as \gls{nrh} values decrease.}
% \agy{6}{due to \om{7}{many} preventive refresh operations}.
\atb{6}{We compare \gls{prac} to} \ous{6}{\param{three} state-of-the-art mitigation mechanisms: 1) Graphene~\cite{park2020graphene}, 2) Hydra~\cite{qureshi2022hydra}, and 3) PARA~\cite{kim2014flipping}.
Our results across 60 different four-core multiprogrammed benign workloads show that \gls{prac} performs 1) better than PARA at \gls{nrh} values lower than 1K and 2) \om{7}{comparably to} Graphene and Hydra at \gls{nrh} values lower than \param{256}, as it performs preventive refreshes \om{7}{less aggressively}, i.e., when a row activation counter gets close to \gls{nrh}.}
\agy{6}{Fourth, we define an \om{9}{availability-oriented} adversarial access pattern that \om{7}{exacerbates} the performance overhead of \gls{prac} to perform a memory performance attack and show that \atb{6}{this} adversarial access pattern \om{7}{1) hogs} up to \ous{12}{\param{94\%}} of DRAM \om{7}{throughput}}
\agy{6}{and \om{7}{2)} reduces system \om{7}{performance} by up to \ous{12}{\param{94.5}}\% (\ous{12}{\param{86.8}}\% on average) across \param{60} workloads.}
\ouscomment{5}{I removed the conclusion for now as we do not introduce the critical data here yet. I expanded our second contribution a bit to give some hindsight about it.
I'm trying to structure in a ``\gls{prac} is bad at high thresholds due to timing parameters, but timely prevention allows it to scale well'' direction in general.}
% \atbcomment{5}{is bandwidth better than availability?}
% \agy{6}{Our simulations show that such} 
% \atb{6}{The} \agy{6}{malicious access pattern can significantly reduce}

% \atb{6}{We compare \gls{prac} to X state-of-the-art XYZ techniques...}\agycomment{6}{Why did you add this? Does this address any of Onur's comments?}
% \ouscomment{6}{I think he did it for consistency after I expanded the 2nd item in conclusions.}

% Therefore, we conclude that \gls{prac} can provide security at low performance and energy overhead, but \ous{6}{is vulnerable to} adversarial memory performance attacks.}
% \agy{2}{across all \ous{4}{workloads}}
% all simulated benign workload mixes
% \agycomment{4}{add a list of contributions}

\ous{4}{We make the following contributions:
\begin{itemize}
[noitemsep,topsep=0pt,parsep=0pt,partopsep=0pt,labelindent=0pt,itemindent=0pt,leftmargin=*]
    \item {We present the first security analysis of \gls{prac} and provide robust \ous{7}{\gls{prac}} configurations against \ous{6}{its} worst-case access pattern.}
    \item {We rigorously evaluate the performance, energy, and cost implications of \ous{4}{\gls{prac}'s different} configurations for modern and future DRAM chips. \om{7}{Our results} show that \ous{6}{\gls{prac} \om{7}{incurs} non-negligible overheads\om{9}{, even for} \om{7}{DRAM chips with \gls{nrh} values higher than 1K, because it increases critical DRAM} timing parameters.}}
    \item {\ous{6}{We compare \gls{prac} to \param{three} \ous{7}{state-of-the-art} mitigation mechanisms for \ous{7}{modern} and future DRAM chips. \ous{7}{Our results show that \gls{prac} 1) underperforms against two of the three mitigation mechanisms for modern DRAM chips \om{8}{with relatively high (i.e., $\geq$1K) \gls{nrh} values} and 2) performs comparably to all three mitigation mechanisms for future DRAM chips with lower \gls{nrh} values, because it performs preventive refreshes in a timely manner.}}}
    \item {We mathematically and empirically \ous{4}{show that} an attacker can exploit \ous{4}{\gls{prac}'s} preventive refreshes to mount memory performance \ous{6}{attacks}~\cite{moscibroda2007memory,mutlu2008parbs,mutlu2007stall,kim2010thread,kim2010atlas,subramanian2016bliss,subramanian2014bliss} \om{7}{and hog a large fraction of DRAM throughput, which in turn, significantly degrades system \om{7}{performance}}.}
    \item {\om{7}{To aid future research in a transparent manner, we open-source our implementations and \om{8}{scripts} at \url{https://github.com/CMU-SAFARI/ramulator2}.\agycomment{7}{Dropped ``for now''}}}
\end{itemize}}
\section{Background \& Motivation}
\label{sec:background}

% This section provides a concise overview of 1)~DRAM organization and operation
% and 2)~DRAM read disturbance.
% For more detail, we refer the reader to prior works {on DRAM and read disturbance}~\cite{ipek2008self,zhang2014half, qureshi2015avatar, liu2012raidr, liu2013experimental, keeth2001dram, mutlu2007stall, moscibroda2007memory, mutlu2008parbs, kim2010atlas, subramanian2014bliss, salp, kim2014flipping,
% hassan2016chargecache, chang2016understanding, lee2017design,  chang2017understanding,  patel2017reaper,kim2018dram, kim2020revisiting, hassan2019crow, frigo2020trrespass, chang2014improving, chang2016low, ghose2018vampire, hassan2017softmc, khan2016parbor, khan2016case, khan2014efficacy, seshadri2015gather, seshadri2017ambit, kim2018solar, kim2019d, patel2019understanding, patel2020beer, lee2013tiered, lee2015decoupled, seshadri2013rowclone, luo2020clrdram, seshadri2019dram, wang2020figaro}.
% \agycomment{4}{Non-SAFARI references: \cite{keeth2001dram, ipek2008self,zhang2014half, qureshi2015avatar}}

% \subsection{DRAM Organization and Operation}
% \label{sec:background_dram}
\head{Organization}
A memory channel connects the processor to a set of DRAM chips, called \agy{4}{\emph{DRAM rank}}.
%\agy{4}{Chips in a DRAM rank operate in lock-step.}
\agy{0}{Each chip has multiple DRAM banks,
% each consisting of multiple DRAM cell arrays {(called \emph{subarrays})} and their local I/O circuitry. {Within a subarray,} 
\agy{2}{in which} DRAM cells are organized as a two-dimensional array {of} rows and columns.}
% DRAM cells in a DRAM row are connected to a wire, called wordline; and cells in a DRAM column are connected to another wire called bitline. 
% {A} DRAM cell {stores one bit of data} {in the form of} electrical charge in {a} capacitor.%, which can be accessed through an access transistor.  
%A wire called \emph{{wordline}} drives the gate of all DRAM cells' access transistors in a DRAM row{. A} wire called {\emph{bitline}} connects all DRAM cells in a DRAM column to a common differential sense amplifier. 
%Therefore, when a wordline is asserted, each DRAM cell in the DRAM row is connected to its corresponding sense amplifier. 
%The set of sense amplifiers in a subarray is called {\emph{the row buffer}}, where the data of {an activated} DRAM row is buffered \agy{4}{to serve a column access.}
% {An access transistor, controlled by the wordline, conducts the cell capacitor to the bitline.}}
% {To access a DRAM row, the row decoder asserts a wire called wordline.}
% {DRAM cells in the same column are connected to a set of differential sense amplifiers called row buffer via a shared wire called bitline.}

% \begin{figure}[h]
%     \centering
%     \includegraphics[width=\linewidth]{figures/background.png}
%     \caption{DRAM organization}
%     \label{fig:dram_organization}
% \end{figure}

\head{Operation}
The memory controller serves memory access requests \ous{6}{with four} \gf{1}{main} DRAM commands\ous{6}{\cite{jedec2020jesd795,jedec2020lpddr5,jedec2015lpddr4, jedec2015hbm,jedecddr,jedec2017ddr4,jedec2012ddr3,kim2012acase,lee2015adaptive,lee2013tiered}}.
%, e.g., row activation ($ACT$), bank precharge ($PRE$), data read ($RD$), data write ($WR$), and refresh ($REF$) \gf{0}{while respecting certain timing parameters to guarantee correct operation~\cite{jedec2020jesd795,jedec2020lpddr5,jedec2015lpddr4, jedec2015hbm,jedecddr,jedec2017ddr4,jedec2012ddr3}.}
\gf{1}{First, the memory controller} issues an $ACT$ command alongside the bank address and row address corresponding to the memory request's address\gf{1}{, which opens (activates) one DRAM row in a DRAM bank.} 
%To read or write data{,} the memory controller first {needs to activate the corresponding row. To do so, it} issues an $ACT$ command alongside the bank address and row address corresponding to the memory request's address. 
%and opens a row corresponding to the memory request's address. 
%When a row is activated, its data is copied \gf{0}{to and temporarily stored at} the row buffer.
%\gf{0}{The latency from the start of a row activation until the data
%is reliably readable{/writable} in the row buffer is called {the} \emph{\gls{trcd}}.}
%\gf{0}{During the row activation process, a DRAM cell loses its charge, and thus, its initial charge needs to be restored ({via} a process called \emph{charge restoration}). 
%The latency from the start of a row activation until the completion of the DRAM cell's charge restoration is called \agy{4}{the \emph{\gls{tras}}}.%DRAM manufacturers provide a built-in safety margin in the nominal timing parameters to account for the worst-case latency in \gls{trcd} and \gls{tras} operations~\cite{}.
%}
\gf{1}{Second, t}he memory controller can read/{write} data from/to \gf{1}{an activated row}  using $RD$/$WR$ commands.
%The changes are propagated to the DRAM cells in the open row. 
%Subsequent accesses to the same row can be served quickly {from the row buffer (i.e., called a \emph{row hit})} without issuing another $ACT$ to the same row. 
% The latency of performing a read/write operation is called \gls{tcl}/\gls{tcwl}. 
\gf{1}{Third, t}o access another row in {an already} activated DRAM bank, the memory controller must issue a $PRE$ command {to} close the opened row {and prepare the bank for a new activation}.
%, creating a \emph{row miss/conflict.} 
%\agy{4}{When the $PRE$ command is issued}, the DRAM chip de-asserts the active row's wordline and precharges the bitlines. The timing parameter for precharge is called {the \emph{\gls{trp}}}.
\agy{2}{The memory controller obeys \ous{6}{many} \gf{1}{timing \agy{2}{constraints} to guarantee correct operation}\ous{6}{\cite{kim2012acase, lee2015adaptive, lee2013tiered, jedec2024jesd795c}}.
Three constraints on minimum delay between commands are 1)~\gls{tras}, from an $ACT$ command to the next $PRE$ command; 2)~\gls{trp}, from a $PRE$ command to the next $ACT$ command; and 3)~\gls{trc}, between two $ACT$ commands as the sum of \gls{tras} and \gls{trp}.}

\head{Refresh}
%{{A} DRAM cell {is} inherently leaky and {loses its} stored electrical charge over time. 
To maintain data integrity, a DRAM cell is periodically refreshed\om{7}{\cite{liu2012raidr, liu2013experimental, chang2014improving}} with a {time interval called \agy{4}{the \emph{\gls{trefw}}}, which is typically} \SI{64}{\milli\second} (e.g.,~\cite{jedec2012ddr3, jedec2017ddr4, micron2014ddr4}) or \SI{32}{\milli\second} (e.g.,~\cite{jedec2015lpddr4, jedec2020jesd795, jedec2020lpddr5}).
%\agy{4}{at normal operating temperature (\om{5}{i.e.}, up to \SI{85}{\celsius}).
%and half of it for the extended temperature range (\om{5}{i.e.}, above \SI{85}{\celsius} up to \SI{95}{\celsius})
%}  
%To {timely} refresh all cells,
\gf{1}{T}he memory controller {periodically} issues a \gls{ref} command with {a time interval called} \agy{4}{the \emph{\gls{trefi}}}, {typically} \SI{7.8}{\micro\second} {(e.g.,~\cite{jedec2012ddr3, jedec2017ddr4, micron2014ddr4}) or} \SI{3.9}{\micro\second} (e.g.,~\cite{jedec2015lpddr4, jedec2020jesd795, jedec2020lpddr5}).
%\agy{4}{at normal operating temperature.} 
When a \ous{6}{rank- or bank-}level refresh\ous{6}{\cite{chang2014improving}} is issued, the DRAM chip internally refreshes several DRAM rows, during which the whole \ous{6}{rank or bank} is busy.
This operation's latency is called \agy{4}{the \emph{\gls{trfc}}}. 
% in DDR4~\cite{jedec2017ddr4}/DDR5~\cite{jedec2020ddr5}.

% \head{DRAM Timing Parameters} 
% \copied{ABACUS}{The memory controller schedules DRAM commands according to certain timing parameters to guarantee correct operation~\cite{jedec2020ddr5,jedec2020lpddr5,jedec2015lpddr4, jedec2015hbm,jedecddr,jedec2017ddr4,jedec2012ddr3}.}
% % {In addition to \gls{trefw}, \gls{trefi}, and \gls{trfc},} three {other} timing parameters are important to understand the rest of the paper: i)~\gls{trc} (Fig.~\ref{fig:dram_organization}b), ii)~\gls{trrd}, iii)~\gls{tras}.}
% \agycomment{0}{Add missing timing constraints.}

\head{Read Disturbance}
\label{sec:background_readdisturbance}
Read disturbance is the phenomenon that reading data from a memory device causes \ous{6}{electrical} disturbance %(e.g., voltage deviation, electron injection, electron trapping) 
on another piece of data that is \emph{not} accessed but physically located nearby the accessed data. Two prime examples of read disturbance in modern DRAM chips are RowHammer~\cite{kim2014flipping} and RowPress~\cite{luo2023rowpress}, where repeatedly \om{7}{activating} (\ous{6}{i.e.,} hammering) or keeping active (\ous{6}{i.e.,} pressing) a DRAM row induces bitflips in physically nearby DRAM rows. In RowHammer and RowPress terminology, \om{7}{a} row that is hammered or pressed is called the \emph{aggressor} row, and the row that experiences bitflips the \emph{victim} row.
{For read disturbance bitflips to occur, 1)~\om{7}{an} aggressor row needs to be activated more than a certain threshold value, \agy{0}{defined as \ous{6}{\gls{nrh}}~\cite{kim2014flipping} {and/}or 2)~\gls{taggon}~\cite{luo2023rowpress} \om{7}{needs} to be large-enough\om{7}{\cite{luo2023rowpress}}. \ous{6}{One way to} avoid read disturbance is to \ous{6}{identify potential aggressor rows and} preventively refresh \ous{6}{their potential} victim rows~\refreshBasedRowHammerDefenseCitations{}.
% \cite{kim2020revisiting, orosa2021deeper, yaglikci2022understanding, luo2023rowpress}
% selectively {throttle} accesses to aggressor rows~\cite{yaglikci2021blockhammer, greenfield2012throttling}, and physically {isolate} potential aggressor and victim rows~\cite{hassan2019crow, konoth2018zebram, saileshwar2022randomized, saxena2022aqua, wi2023shadow, woo2023scalable}. 
%These \agy{4}{solutions} aim to perform preventive actions before the cumulative effect of an aggressor row's \emph{activation count} and \emph{on time} causes read disturbance bitflips.
}}

\head{\agy{2}{\gls{prac}}}
\ous{6}{Various prior works discuss the use of per-row activation counters to detect how many times each row in DRAM is activated within a refresh interval~\cite{kim2014flipping,kim2014architectural,bennett2021panopticon,kim2023ddr5,yaglikci2021security}.}
\agy{2}{A recent update \ous{6}{(as of \agy{6}{April} 2024) of} \gf{2}{the JEDEC} DDR5 specification~\cite{saroiu2024ddr5, jedec2024jesd795c} introduces a \ous{6}{similar} on-DRAM-die read disturbance mitigation mechanism called \gls{prac}
\agy{2}{(explained in \secref{sec:briefsummary}), which}
aims to ensure robust operation at low overhead by preventively refreshing victim rows when necessary.
\agy{2}{Although \gls{prac} is a promising DRAM specification advancement,}
\emph{no} prior work rigorously analyzes \agy{2}{\gls{prac}'s impact on security, performance, energy, and cost for} modern and future systems.}
\section{A Brief Summary of RFM and PRAC}
\label{sec:briefsummary}

\agy{2}{This section briefly explains the \gls{rfm} command, \gls{prac} mechanism, and assumptions \ous{6}{we use} for \gf{2}{our} \ous{6}{evaluations}.}

\head{RFM \gf{1}{C}ommand}
\gls{rfm} is a DRAM command that provides the DRAM chip with a time window (e.g.,~\param{\SI{195}{\nano\second}}~\cite{jedec2024jesd795c}) so that the DRAM chip preventively refreshes potential victim rows.
The DRAM chip is responsible for identifying and preventively refreshing potential victim rows, and the memory controller is responsible for issuing \gls{rfm} commands.

\head{\gls{prac} Overview}
\gls{prac} implements an activation counter for each DRAM row, and thus accurately measures the activation counts of \emph{all} rows.
When a row's activation count reaches a threshold, \ous{6}{the DRAM chip} asserts a back-off signal \ous{6}{which forces} the memory controller to issue an RFM command.
\ous{6}{The DRAM chip} preventively refreshes potential victim rows upon receiving an RFM command. 

\head{Assumptions about the \gls{prac} Mechanism}
We make two assumptions:
1)~\gls{prac} always refreshes potential victims of the row with the maximum activation count \ous{7}{during each \gls{rfm} command (even if the maximum activation count is \emph{not} close to \gls{nrh})}\footnote{\ous{9}{The specification does \emph{not} enforce refreshing the victims of \om{10}{the} aggressor with \om{10}{the maximum} activation count~\cite{jedec2024jesd795c}.}} and
2)~\ous{6}{physically-adjacent DRAM rows} can experience bitflips when a DRAM row is \ous{7}{activated} more than a threshold value, denoted as \gls{nrh}.

% \head{RFM-Only Mechanism}
% As described in the updated and early DDR5 specifications~\cite{jedec2020jesd795}, RFM-only mechanism issues an \gls{rfm} command when the total number of activations targeting a bank exceeds a predefined threshold value \gls{rfmth}.
% \agycomment{1}{is this RFMth?} \ouscomment{1}{It is RAAIMT. But Mithril uses RFMth and I think it is a lot clearer than. I can update it or add a footnote perhaps.}
% RFM-only mechanism does \emph{not} implement any back-off signal from the DRAM chip to the memory controller.

% \head{RFM-With-Backoff Mechanism}
% As described in the DDR5 update~\cite{saroiu2024ddr5, jedec2024jesd795c}, the DRAM chips can implement a mechanism called \gls{prac}, which implements an activation counter per DRAM row for accurately tracking hammer counts and a back-off protocol which informs the memory controller that the DRAM chip needs an \gls{rfm} command to preventively refresh victim rows. The novelty of this update is the back-off protocol compared to the RFM-only mechanism. Therefore, we call this mechanism \emph{RFM-with-backoff}.
% With respect to the update released in April, 2024~\cite{saroiu2024ddr5, jedec2024jesd795c}. 

\head{\gls{prac}'s \gf{1}{O}peration and \gf{1}{P}arameters}\omcomment{5}{Hard to follow}
\gls{prac} increments the activation count of a DRAM row while the row is \ous{6}{being closed (i.e., during precharge)}, which increases \gls{trp} and \gls{trc}~\cite{jedec2024jesd795c}).\footnote{\ous{7}{When \gls{prac} is enabled, activation counters are incremented with internal reads and writes before a row is closed.
The counter update causes a delay between the time of receiving a precharge command and \emph{actually} precharging the row.
Because of this delay\om{8}{,} 1) $\trp$ increases by \param{\SI{21}{\nano\second}} \ous{8}{(\param{+140\%})} and 2) $\tras{}$, $\trtp{}$, and $\twr{}$ reduces by \param{\SI{16}{\nano\second}} \ous{8}{(\param{-50\%})}, \param{\SI{2.5}{\nano\second}} \ous{8}{(\param{-33\%})}, and \param{\SI{20}{\nano\second}} \ous{8}{(\param{-66\%})}~\cite{jedec2024jesd795c}.
Combined effect of these timing parameters result in a $\trc$ increase of \SI{5}{\nano\second} \ous{8}{(\param{+10\%})} (for DDR5-3200AN speed bin~\cite{jedec2024jesd795c}).}}
The DRAM chip asserts the back-off signal when a row's activation count reaches a fraction of \gls{nrh}, denoted as \gls{aboth}, where the fraction can be configured to either 70\%, 80\%, 90\%, \ous{4}{or 100\%}~\cite{jedec2024jesd795c}.
The memory controller receives the back-off signal \ous{6}{between the time after \ous{7}{a} command \ous{7}{that closes rows (e.g., precharge or refresh)} is issued and a small latency after \ous{7}{the same} command's completion} (e.g., \param{$\approx$\SI{5}{\nano\second}}~\cite{jedec2024jesd795c}).
\ous{6}{The memory controller and the DRAM chip go through three phases when the back-off signal is asserted.}
\ous{6}{First, \om{7}{during \gls{taboact}}\ous{6}{\cite{jedec2024jesd795c}}, the memory controller has a limited time window (e.g., \param{\SI{180}{\nano\second}}~\cite{jedec2024jesd795c}) to serve requests after receiving the back-off signal.}
\agy{3}{A} DRAM row can receive up to \ous{2}{\gls{taboact}}/\gls{trc} activations \ous{6}{in this window.}
\ous{6}{Second, \om{7}{during} the \emph{recovery period}\ous{6}{\cite{jedec2024jesd795c}}, the memory controller issues a number of \gls{rfm} commands, which we denote as $\bonrefs{}$ (e.g., 1, 2 or 4~\cite{jedec2024jesd795c}).}
\agy{3}{An \gls{rfm} command} can further increment the activation count of a row before its potential victims are refreshed.
\ous{6}{Third, \om{7}{during} the \emph{delay period} or \gls{tbodelay}\ous{6}{\cite{jedec2024jesd795c}}, the DRAM chip \emph{cannot} reassert the back-off signal until it receives a number of \gls{act} commands, which we denote as $\bonacts{}$  (e.g., 1, 2 or 4~\cite{jedec2024jesd795c}).}\footnote{\agy{4}{Current DDR5 specification~\cite{jedec2024jesd795c} \atb{6}{notes} that $\bonrefs$ and $\bonacts$ always \atb{6}{have} the same value. \atb{6}{To comprehensively assess PRAC's security guarantees, we use different values for the two parameters \emph{only} in our security analysis of PRAC (\secref{sec:configurationandsecurity}).}}}
% \ous{6}{%which we separated for analysis clarity. 
% \atb{6}{Unless otherwise stated, the number following the hyphen after PRAC} specifies the value of $\bonacts{}$ and $\bonrefs$ in the mechanism name, e.g., \gls{prac}-4 denotes $\bonacts{}$ and $\bonrefs{}$ is equal to 4}.}}
% The number of refreshes within a recovery period (parameter N of \gls{prac}-N), and the number of activates within a delay period is defined to be same in the JEDEC's DDR5 specification~\cite{jedec2024jesd795c}. In our analysis, to avoid confusion we separate these parameters as $\bonrefs$ and $\bonacts$, respectively.
% \agy{3}{after asserting the back-off signal.}
% The memory controller has a limited time window (e.g., \param{\SI{180}{\nano\second}}~\cite{jedec2024jesd795c}), denoted as \ous{2}{\gls{taboact}}, to serve requests after receiving the back-off signal and before \ous{2}{entering the \emph{recovery period} \agy{3}{where the memory controller issues 
% \agy{4}{a number of} \gls{rfm} commands, \agy{4}{which we denote as} \ous{3}{$\bonrefs{}$ (e.g., 1, 2 or 4~\cite{jedec2024jesd795c})}.}}
% \agy{4}{The time window until receiving $\bonacts{}$ row activations is called}
% the \emph{delay period}, or \gls{tbodelay}.
% based on the refresh configuration, 
\agy{3}{Considering these \ous{6}{three phases}, \secref{sec:configurationandsecurity} calculates} the highest \om{7}{achievable} activation count \om{7}{to any DRAM row} \agy{3}{in a \gls{prac}-protected system.}

\head{\om{7}{\gls{rfm} and} \gls{prac} \gf{1}{I}mplementations}
We analyze four different \om{7}{\gls{rfm} and} \gls{prac} implementations:
% of in-DRAM read disturbance mitigation mechanisms: 
1)~\ous{6}{\emph{\gls{prfm}}}, where the memory controller issues an \gls{rfm} command \emph{periodically} when the total number of activations \ous{6}{to} a bank reaches a predefined threshold value called \om{7}{\gls{rfmth}} with \emph{no} back-off signal from the DRAM chip, as described in early DDR5 standards~\cite{jedec2020jesd795};
2)~\ous{6}{\emph{\gls{prac}-N}}, where the memory controller issues N back-to-back \gls{rfm} commands \emph{only} after receiving a back-off signal from the DRAM chip, as described in the \gf{2}{latest JEDEC} DDR5 standard~\cite{saroiu2024ddr5, jedec2024jesd795c};\ouscomment{8}{reordered}
3)~\ous{6}{\emph{\gls{prac}+\gls{prfm}}}, where the memory controller issues an \gls{rfm} command $i$)~\ous{6}{when the total number of activations to a bank reaches \gls{rfmth}} \ous{7}{or} $ii$)~\om{7}{it receives} a back-off signal from the DRAM chip.
\ous{7}{\ous{8}{PRAC-N}\ouscomment{8}{-optimistic is undefined, dropped it} \om{8}{implementations} are \emph{not} secure at \gls{nrh} values lower than \param{20}.
Therefore, combining \gls{prac} and \gls{prfm} enables security at lower \gls{nrh} values at the cost of potentially refreshing \ous{8}{the victims of aggressor rows} \om{8}{whose activation counts} are \emph{not} close to \gls{nrh}; and}
4)~\ous{6}{\emph{\om{7}{\gls{prac}-Optimistic}}, which \om{7}{is the same as} \ous{7}{the default \gls{prac} configuration advised by JEDEC~\cite{jedec2024jesd795c} (i.e., \gls{prac}-4)} \emph{without} \om{8}{any change in timing parameters (including \gls{trp} and \gls{trc}~\cite{jedec2024jesd795c})}.}
\section{Adversarial Access Pattern: The Wave Attack}
\label{sec:adversarial}

\head{Threat Model}
To account for the worst case, we assume that the attacker
1)~knows the physical layout of DRAM rows \atb{6}{(as in~\cite{yaglikci2021blockhammer})},
2)~accurately detects when a row is internally refreshed (preventively or periodically \atb{6}{as in U-TRR~\cite{hassan2021utrr}}), \ous{6}{and}
3)~precisely times \emph{all} DRAM commands \ous{6}{except} \gls{ref} and \gls{rfm} commands \atb{6}{(as in~\cite{yaglikci2021blockhammer,hassan2021utrr})}. 

\head{Overview}
The adversarial access pattern aims to achieve the highest number activation count for a given row in a \gls{prac}-protected DRAM chip by overwhelming \gls{prac} \ous{6}{using} a number of decoy rows, similar to the 
% adversarial access patterns in prior works, referred to as 
\emph{wave attack}~\cite{yaglikci2021security, devaux2021method}.\ouscomment{6}{removed fading attack}
In this access pattern, the attacker hammers a number of rows in a balanced way, such that \gls{prac} can perform preventive refreshes \emph{only} for a \ous{6}{small} subset of the hammered rows when an \gls{rfm} is issued. When an aggressor row's victims are refreshed, the attacker excludes the aggressor row in the next round of activations. By doing so, this adversarial access pattern achieves the highest possible \ous{6}{activation count} for the row whose victims are preventively refreshed latest.
\section{Configuration of PRAC and Security Analysis}
\label{sec:configurationandsecurity}
\agy{1}{This section investigates \om{7}{different \gls{rfm} and} \gls{prac} configurations and their impact on security under the wave attack.}

\head{Notation} We denote the set of rows that the wave attack hammers in round $i$ as $R_i$ and the \ous{6}{number of rows in} $R_i$ as $|R_{i}|$.

\head{Key Parameters} We assume
a \ous{6}{\emph{blast radius}} of 2~\cite{kogler2022half},
a \gls{trc} of \SI{52}{\nano\second}~\cite{jedec2024jesd795c},
and a \gls{trfm} of \SI{350}{\nano\second}~\cite{jedec2024jesd795c}, which allows an RFM command to refresh four victim rows of one aggressor row.

\head{PRFM}
In round 1, the wave attack \agy{1}{hammers each row in $R_1$ once, causing the memory controller to issue $\lfloor(|R_{1}|/\rfmth{})\rfloor$ \gls{rfm} commands, \ous{5}{each refreshing} the \ous{6}{four} victims of one aggressor row.}
% each of which
% starts by \agy{1}{hammering} a set of rows $R_0$ with no assumed activations.
% In the first step, each row in $R_0$ is activated once and the memory controller issues $\lfloor \frac{|R_0|}{\rfmth{}} \rfloor$ RFM commands, refreshing a portion of these rows.
\agy{1}{In round 2, the wave attack continues hammering} the non-refreshed rows $R_2$, where $|R_2| = |R_1| - \lfloor ({|R_1|}/{\rfmth{}}) \rfloor$.
\agy{1}{By repeating this calculation $i$ times, \atb{6}{Equation}~\ref{eqn:rfmrec} evaluates the} number of rows \ous{7}{with victims that are not refreshed} at \agy{1}{an arbitrary round $i$} ($|R_{i}|$). 

\vspace{-0.5em}
\agycomment{3}{double check. Is ``$n\times$'' stale?}
\ouscomment{3}{yes i removed it now. we assume it refreshes the victims of one aggressor, which is outlined in key parameters above}
\begin{equation}
\label{eqn:rfmrec}
|R_{i}| = |R_{1}| - 
\Bigl\lfloor
\frac{
\sum_{k=1}^{i-1} |R_{k}|
}{
RFM_{th}
}
\Bigr\rfloor
\end{equation}

\ous{6}{To cause bitflips,} \agy{1}{the wave attack \ous{6}{must} make sure that \ous{3}{1)} at least one aggressor row's victims are \emph{not} refreshed \ous{1}{by an RFM command} at round $N_{RH}$, i.e., $|R_{N_{RH}}| > 0$, \ous{1}{and \ous{3}{2)} the time taken by the attacker's row activations and RFM preventive refreshes do \agy{2}{\emph{not}}\agycomment{2}{Always \emph{emph} negatives} exceed \gls{trefw}, i.e., aggressor's victims are \agy{2}{\emph{not}} periodically refreshed} \om{7}{before being activated \gls{nrh} times}.
We rigorously sweep \om{7}{the} \agy{5}{wave attack's} configuration parameters and identify the maximum hammer count \ous{6}{of} an aggressor row \ous{6}{before its victims are refreshed}.}
% For an \gls{rfm}-only system to be secure against RowHammer bitflips it should not be possible for an attacker to obtain $|R_{N_{RH}-1}| > 0$ within \gls{trefw} for any $R_0$.
% Based on this observation, we perform an analysis of the maximum steps possible under a system with \gls{rfm} protection for varying \gls{rfmth} configurations and initial row set sizes.

\head{\gls{prac}-N}
\agy{1}{We adapt \ous{2}{our} \om{7}{\gls{prfm}} wave attack \ous{2}{analysis} to \gls{prac}-N by leveraging \ous{1}{two} key insight\ous{1}{s}:}
\ous{2}{First,} \gls{prac}-N mechanism will \emph{not} preventively refresh any row until a row's activation count reaches \glsfirst{aboth}.
\ous{2}{We \ous{3}{prepare} rows in $R_1$ \ous{3}{such that} each row is already hammered \gls{aboth}-1 times.
Doing so, the number of rounds necessary to induce a bitflip is reduced by \gls{aboth}-1.}
\ous{2}{Second, \agy{3}{at least one row's activation counter remains above \gls{aboth} across \emph{all} rounds after initialization until the end of the wave attack.}}
This causes \gls{prac}-N to \agy{3}{assert the back-off signal as frequently as possible, i.e., with a time period containing a recovery period ($\bonrefs\times\trfm$), a delay period (\gls{tbodelay}), and a window of normal traffic (\gls{taboact}).}
\agy{2}{Leveraging these insights,} \ous{2}{we update \atb{6}{Equation}~\ref{eqn:rfmrec} to \agy{3}{derive} \atb{6}{Equation}~\ref{eqn:pracrec}\agy{3}{.}}
\agycomment{1}{you do not initialize rows. You do initialize activation counters of the rows}
% activation counter

\vspace{-0.5em}
\begin{equation}
\label{eqn:pracrec}
|R_{i}| = |R_{1}| - \bonrefs \times
\Bigl\lfloor
\frac{
\sum_{k=1}^{i-1} |R_{k}|
}{
\bonacts + (\taboact / \trc)
}
\Bigr\rfloor
\end{equation}

For a \agy{1}{\gls{prac}-N} system to be secure\agy{5}{,}
an attacker \ous{6}{should \emph{not} be able} to obtain $|R_{\nrh{}-\aboth{}}| > 0$ within \gls{trefw} for any $R_1$.\omcomment{5}{Sounds cryptic. Expand in extended version and provide insight.}
We \agy{5}{analyze} the maximum \ous{7}{hammer count of an aggressor row before its victims are refreshed} \agy{5}{in a \gls{prac}-N-protected system} 
% under a system with \gls{prac}-N protection 
for \om{7}{a wide set of} \gls{aboth} and \agy{5}{$|R_{1}|$ configurations}. 
% and initial row set sizes.

\begin{figure}[h]
\centering
\includegraphics[width=\linewidth]{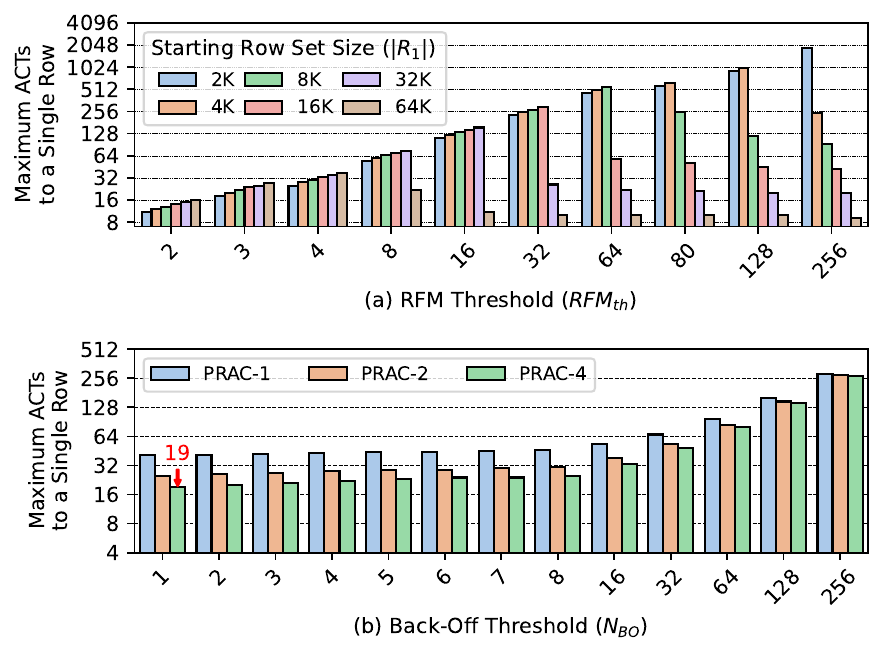}
\caption{Maximum activations \atb{6}{to} a row \atb{6}{allowed by} \ous{4}{(a) \om{7}{\gls{prfm}} and (b) \om{7}{\gls{prac}-N}}}
\label{fig:rfmpracanalysis}
\end{figure}

\head{Configuration Sweep}
\agy{5}{\figref{fig:rfmpracanalysis} shows the maximum activation count an aggressor row can reach before its victims are refreshed (y-axis) for \gls{prfm} and \om{7}{\gls{prac}-N} in \figsref{fig:rfmpracanalysis}a and~\ref{fig:rfmpracanalysis}b, respectively.
\figref{fig:rfmpracanalysis}a shows \ous{6}{the bank activation threshold to issue an RFM command ($\rfmth{}$)} on the x-axis and starting row set size ($|R1|$) color-coded. \figref{fig:rfmpracanalysis}b shows \ous{6}{the back-off threshold ($\aboth{}$)} on the x-axis and $\bonrefs$ color-coded.}
% Figure~\ref{fig:rfmpracanalysis}\ous{4}{a} \agy{1}{shows} our analysis where
% \agy{1}{x- and y-axes show \gls{rfmth} and the maximum activation count an aggressor row can reach before its victims are refreshed, respectively. Different colors represent different starting row set sizes $|R_{1}|$.}
% the x-axis shows the different $\rfmth{}$ values, the y-axis shows the maximum possible number of activations on a row before it is refreshed, and each bar identifies a different starting row set size ($|R_{0}|$).

\ous{4}{From \agy{5}{\figref{fig:rfmpracanalysis}a}, we observe that} \agy{1}{to prevent bitflips for very low \gls{nrh} values (e.g., 32 \om{7}{on the y axis}), \gls{rfmth} should be configured to very low values (e.g., $<$4), as only such \gls{rfmth} values results in activation counts less than \gls{nrh} for all $|R_1|$ values}.
% \agycomment{5}{Do we have only one observation now?}
\ous{2}{From \agy{5}{\figref{fig:rfmpracanalysis}b}, we observe that \gls{prac}-N provides security at \gls{nrh} values as low as \ous{12}{\param{20}} \atb{6}{(because a row can receive \ous{12}{\param{19}} activations \om{7}{as annotated})} when configured to \agy{5}{1)~}trigger a back-off \ous{6}{as frequently as possible ($N_{BO}=1$)} and \agy{5}{2)~issue \param{four} RFMs}
% \param{four} aggressors
in the recovery period (i.e., \gls{prac}-4).}
\ous{10}{For the remainder of our study, we assume we can accurately determine \gls{nrh} and configure \gls{prfm} and \gls{prac} protected systems to avoid all bitflips using these secure thresholds (which is a difficult problem in itself, given that determining \gls{nrh} for every row is not easy, as shown by multiple works~\cite{kim2014flipping, orosa2021deeper, luo2023rowpress, kim2020revisiting, saroiu2022configure, olgun2023understanding, zhou2023threshold}).}
% Second, the \param{three} quantized back-off trigger percentages (i.e., 70\%, 80\%, 90\%) does \emph{not} provide security against read disturbance bitflips when $\aboth{} = \nrh{}$ at $\nrh{} < 32$.
% However, \gls{aboth} can be configured as lower a value than \gls{nrh} to reach the desired maximum activation guarantees against read disturbance attacks.
% Based on these \param{N} observations we conclude that \gls{prac}-N provides secure execution against read disturbance bitflips at very low \gls{nrh} values.}
% \agycomment{1}{Figure does not have x- and y-axes}
% \agycomment{1}{Let's put the percentages on the top left corner within each subplot}
\agycomment{2}{Does it make sense to convert this into a scatter plot with a y=x line, separating safe and unsafe configurations? I am not sure if it makes sense because it is not clear how x-axis is affiliated with NRH}
\agycomment{2}{What about \gls{prac}-\gls{rfm}?}
\section{Experimental \agy{1}{Evaluation}}
\label{sec:methodology}

\agy{1}{We evaluate \gls{prac}'s overheads on system performance, DRAM energy consumption, and DRAM chip area for \ous{6}{existing} and future DRAM chips, by sweeping \gls{nrh} from \param{1K} down to \ous{12}{\param{20}}.
We compare \gls{prac}'s overheads against \param{three} read disturbance mitigation mechanisms: 
1)~Graphene~\cite{park2020graphene}, the state-of-the-art \gf{5}{mechanism} that maintains row activation counters completely within the processor chip; 2)~PARA~\cite{kim2014flipping}, the state-of-the-art mechanism that does \emph{not} maintain any counters; and
3)~Hydra~\cite{qureshi2022hydra}, the state-of-the-art mechanism that maintains counters in the DRAM chip and cache\atb{6}{s} them in the processor chip.}
\agy{1}{To evaluate performance and DRAM energy consumption, we conduct cycle-level simulations using Ramulator~\gf{5}{2.0}~\cite{ramulator2github, luo2023ramulator2}, integrated with DRAMPower~\cite{drampower}.}
We extend Ramulator~\gf{5}{2.0}~\cite{ramulator2github, luo2023ramulator2} with the implementations of \gls{prac}, \gls{rfm}, \agy{1}{and the back-off signal \ous{8}{(as specified in the latest JEDEC DDR5 DRAM specification as of April 2024~\cite{jedec2024jesd795c})}.}
\agy{2}{We evaluate system \ous{7}{performance} using the weighted speedup metric\om{7}{\cite{eyerman2008systemlevel, snavely2000symbiotic}}.} 
% \gls{abo} alongside Graphene~\cite{park2020graphene} and PARA~\cite{kim2014flipping}.

% Copied from BreakHammer
\tabref{table:system_configuration} shows our system configuration.
We assume a realistic quad-core system, connected to a dual-rank memory with \param{eight} bank groups, each containing \param{four} banks (\param{64} banks in total). The memory controller employs the \ous{1}{FR-FCFS \gf{5}{memory scheduler}\om{7}{\cite{frfcfs, zuravleff1997controller}} with a Cap on Column-Over-Row Reordering (FR-FCFS+Cap) of \param{four}~\cite{mutlu2007stall}}.\agycomment{2}{there should be a paper which propose -Cap for the first time. We should cite it.}
\ous{2}{We extend} the memory controller to \ous{1}{delay} the requests that \emph{cannot} be served within \gls{taboact}.
\agycomment{1}{These two are different schedulers. Check how they are mentioned in old papers~\cite{mutlu2007stall, subramanian2014bliss, subramanian2015application, mutlu2008parbs}}
\ouscomment{1}{I'm not sure if the citation is enough. Old papers only cite FRFCFS.}
\agycomment{1}{eliminate? I hope this is not real.}
\ouscomment{1}{ignores while making scheduling decisions, changed to delay}

\newcolumntype{C}[1]{>{\let\newline\\\arraybackslash\hspace{0pt}}m{#1}}
\begin{table}[ht]
\scriptsize
\centering
\caption{{Simulated} System Configuration}
\begin{tabular}{l|C{5.8cm}}
 \hline
 \textbf{Processor} & {\SI{4.2}{\giga\hertz}, 4-core, 4-wide issue, {128-entry} instr. window} \\ \hline
 \textbf{Last-Level Cache} & {64-byte} cache line, 8-way {set-associative, \SI{8}{\mega\byte}} \\ \hline
 \textbf{Memory Controller} & {64-entry read/write request queues; Scheduling policy: FR-FCFS+Cap of \param{4}\om{7}{\cite{mutlu2007stall}}}; Address mapping: MOP~\cite{kaseridis2011minimalistic} \\ \hline
 \textbf{Main Memory} & DDR5 \om{7}{DRAM~\cite{ramulator2github}}, 1 channel, 2 ranks, 8 bank groups, 4 banks/bank group, {6\atb{6}{4}K} rows/bank\\ \hline
 \end{tabular}
 \label{table:system_configuration}
\end{table}

% Copied from BreakHammer
\head{Workloads}
We evaluate applications from five benchmark suites: SPEC CPU2006~\cite{spec2006}, SPEC CPU2017~\cite{spec2017}, TPC~\cite{tpc}, MediaBench~\cite{fritts2009media}, and YCSB~\cite{ycsb}. We group all applications into three memory\om{7}{-}intensity groups based on their \ous{1}{row buffer} misses per kilo instructions (\ous{1}{RBMPKIs}), similar to prior works~\cite{olgun2024abacus, bostanci2024comet}. These groups are High (H), Medium (M), and Low (L) for the lowest MPKI values of \ous{1}{\param{10, 2, and 0}}\agycomment{2}{0 is not necessary you have three regions split by two thresholds}, respectively. Then, we create \ous{1}{\param{60}} workload mixes \ous{1}{with \param{10} of each HHHH, MMMM, LLLL, HHMM, MMLL, and LLHH combination types}. \ous{1}{We simulate each workload mix until all cores execute \param{100}M instructions \ous{2}{or \param{5} billion cycles}.}
\agycomment{2}{should we mention the hard stop at N billion cycles?}
\agycomment{1}{We mentioned H, M, and L workload types but never talked about how they are combined in mixes.}

% \agycomment{1}{double check}
% \agycomment{1}{don't forget to update these thresholds}
\subsection{\ous{7}{Performance Evaluation}}
\label{sec:evaluation}
\label{subsec:perfeval}
\ouscomment{7}{Storage Evaluation is a separate subsection now}

\begin{figure}[h]
\centering
\includegraphics[width=\linewidth]{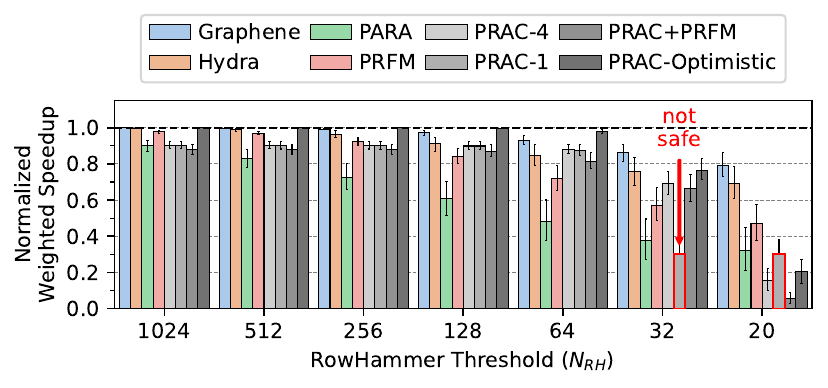}
\caption{Performance impact of evaluated \agy{2}{read disturbance} mitigation mechanisms on \ous{6}{60 benign} four-core workloads}
\label{fig:benign_scaling}
\end{figure}

\figref{fig:benign_scaling} presents the performance overheads of the evaluated \agy{2}{read disturbance} mitigation mechanisms as \gls{nrh} decreases.
\om{8}{Axes respectively} \agy{5}{show \ous{8}{the} \gls{nrh} values} \om{8}{(x axis)} \ous{7}{and system performance \om{8}{(y axis)} in terms of weighted speedup~\cite{eyerman2008systemlevel, snavely2000symbiotic} normalized to} \ous{6}{a} baseline with \emph{no} read disturbance mitigation \om{8}{(higher y value is better)}. \om{8}{Different} \ous{7}{bars \om{9}{identify} different} read disturbance mitigation mechanisms \ous{12}{and red edge color indicates read disturbance vulnerable configurations}.

We make \param{\ous{9}{nine}} observations from \figref{fig:benign_scaling}.\ouscomment{9}{1st paragraph: increased timings}
\ous{7}{First, \om{8}{as \gls{nrh} decreases,} performance overheads of all studied mitigation mechanisms increases, \om{10}{as expected, due to the more frequent mitigating actions (i.e., preventive refreshes) performed}.}

\om{10}{\head{Effect of Increased Timing Parameters}} 
\ous{7}{Second}, \om{8}{at an \gls{nrh} of 1K,} \ous{6}{\gls{prac}-4 performs similarly to PARA and is outperformed by Graphene, Hydra, and \gls{prfm} \om{8}{averaged} across all workloads.
We attribute \om{8}{\gls{prac}'s} non-negligible overhead at \om{7}{relatively} high \gls{nrh} values \om{7}{(i.e., \ous{12}{\param{9.9}}\% average and \ous{12}{\param{13.1}}\% maximum)} to \gls{prac}'s increased DRAM timing parameters~\cite{jedec2024jesd795c} \om{7}{(see \secref{sec:briefsummary})} as \ous{7}{\gls{prac}-Optimistic} (i.e., \ous{7}{\gls{prac}-4} \emph{without} increased DRAM timing parameters) \om{7}{leads to} an average (maximum) \om{7}{system} \ous{7}{performance overhead} of \om{8}{only} \ous{12}{\param{0.002}}\% (\ous{12}{\param{0.01}}\%) across the same workloads at the same 1K threshold.}
\ous{9}{Third}, \ous{9}{between} \gls{nrh} values \ous{9}{of 1K and \ous{12}{64}}, \gls{prac}-Optimistic outperforms all evaluated mitigation mechanisms, \om{10}{demonstrating that \gls{prac} without increased timing parameters has good potential}.

\om{10}{\head{\gls{prac} Scales Well Until \ous{12}{$\bm{\nrh{} = 64}$}}}
\ous{9}{Fourth}, \om{8}{when \gls{nrh} decreases from 1K to \ous{12}{64},} \ous{7}{\gls{prac}-4's average (maximum) system performance overhead across all workloads increases from \ous{12}{\param{9.9}}\% (\ous{12}{\param{13.1}}\%) to \ous{12}{\param{11.8}}\% (\ous{12}{\param{15.7}}\%).
In contrast, Graphene and Hydra's average (maximum) system performance \om{9}{overheads} across all workloads increase from \param{0.03}\% (\param{0.1}\%) and \ous{12}{\param{0.2}}\% (\ous{12}{\param{1.1}}\%) to \ous{12}{\param{6.9}}\% (\ous{12}{\param{13.0}}\%) and \ous{12}{\param{15.4}}\% (\ous{12}{\param{25.9}}\%), respectively.
\ous{9}{Fifth}, between \gls{nrh} values of 128 and \ous{12}{64}, \gls{prac}-4 outperforms Hydra and performs similarly to Graphene.
We attribute the relative improvement against Graphene and Hydra as \gls{nrh} decreases to \om{9}{\gls{prac}'s more} accurate tracking \om{9}{of} aggressor row \om{9}{activations} and \om{9}{the resulting less aggressive} preventive \om{10}{refreshes performed}.

\om{10}{\head{\gls{prac} Overheads Shoot Up at \ous{12}{$\bm{\nrh{} \leq 64}$}}}
\ous{9}{Sixth}, \om{8}{between \gls{nrh} values of} \ous{12}{64} to \ous{12}{20}, \gls{prac}-4 and \gls{prac}-Optimistic's average (maximum) system performance overheads across all workloads significantly increase from \ous{12}{\param{11.8}}\% (\ous{12}{\param{15.7}}\%) and \ous{12}{\param{2.3}}\% (\ous{12}{\param{4.7}}\%) to \ous{12}{\param{84.7}}\% (\ous{12}{\param{94.0}}\%) and \ous{12}{\param{79.6}}\% (\ous{12}{\param{90.8}}\%), respectively.
\ous{9}{We attribute the significant increase in system performance overhead to \gls{prac} performing more frequent preventive refreshes.}
For example, with \gls{prac}-4 \ous{8}{at \gls{nrh} values of \ous{12}{64 and 20}}, the four-core benign workload of \emph{523.xalancbmk}, \emph{435.gromacs}, \emph{459.GemsFDTD}, and \emph{434.zeusmp} trigger \ous{12}{11.9} and \ous{12}{129.0} recovery periods \ous{8}{per million cycles}, resulting in \ous{12}{11.2}\% and \ous{12}{87.7}\% system performance overhead, respectively.
\ous{9}{Seventh}, at an \gls{nrh} of \ous{12}{20}, Graphene and Hydra outperform \emph{all} evaluated \gls{prac} \ous{8}{and \gls{prfm} implementations}.}

\om{10}{\head{\gls{prfm} Performs Poorly}}
\ous{9}{Eighth}, \ous{12}{when \gls{nrh} decreases from 1K to 20}, the average (maximum) \ous{7}{system performance overhead} of \gls{prfm} increases from \ous{12}{\param{2.1}\% (\ous{12}{\param{4.0}}\%) to \ous{12}{\param{53.0}}\% (\ous{12}{\param{68.7}}\%).}
\ous{6}{We attribute this significant overhead increase to \gls{prfm}'s configuration against \om{7}{the} wave attack \ous{7}{drastically increasing} the frequency of \ous{7}{preventive refreshes \om{7}{as \gls{nrh} decreases,} similar to \gls{prac}} \om{10}{for \gls{nrh} between \ous{12}{64 and 20}}.}
\ous{9}{Ninth}, pairing \gls{prac}\ous{2}{-4} with \agy{2}{\gls{prfm}} \ous{1}{increases \gls{prac}'s \ous{7}{system performance overhead}} by an average (maximum) of \ous{12}{\param{24.4}}\% (\ous{12}{\param{66.1}}\%) \ous{2}{across all $\nrh{}$ values}.
\ous{1}{\agy{2}{This is because} \gls{prac}'s secure configurations (\secref{sec:configurationandsecurity}) already preventively refresh all rows before they reach a critical level.
Therefore, pairing \gls{prac}'s secure configurations with \agy{2}{\gls{prfm} causes performance degradation due to unnecessary \ous{7}{preventive refreshes}.}}

\om{8}{We} \ous{6}{conclude that
1)~\gls{prac}'s increased DRAM timing parameters incur significant overheads even under infrequent \ous{7}{preventive refreshes for \ous{8}{modern} DRAM chips \ous{8}{(i.e., \gls{nrh} $=$ 1K)}},
2)~\gls{prac} shows \ous{7}{similar} performance \ous{6}{to Graphene and outperforms Hydra} \ous{7}{as \gls{nrh} values decrease (until \ous{12}{\param{32}}, where \gls{prac} starts performing significantly worse)},}
3)~\gls{prfm} does \emph{not} scale well \ous{6}{with decreasing \gls{nrh} values} and incurs significant \ous{7}{system performance} loss, and
4)~pairing \gls{prac} with \gls{prfm} \om{10}{provides} no \ous{7}{system performance} advantage.

\subsection{Energy \agy{1}{Evaluation}}

\begin{figure}[h]
\centering
\includegraphics[width=\linewidth]{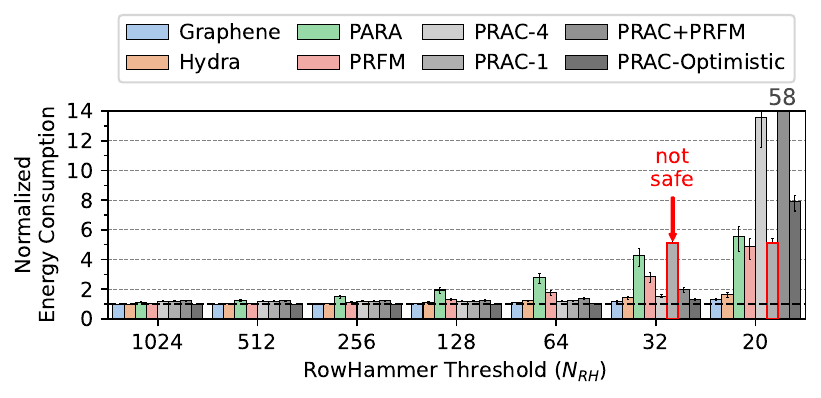}
\caption{Energy impact of evaluated read disturbance mitigation mechanisms on \om{7}{\param{60} benign} four-core workloads}
\label{fig:benign_energy}
\end{figure}

\agycomment{2}{Removed the first sentence as it is redundant with the methodology.}
\figref{fig:benign_energy} presents the energy consumption of the evaluated \agy{2}{read disturbance} mitigation mechanisms \agy{2}{(y-axis)} as \gls{nrh} decreases \agy{2}{(x-axis).
Energy consumption is normalized to \ous{7}{a} baseline with \emph{no} read disturbance mitigation.} 

We make \param{\ous{7}{four}} observations from \figref{fig:benign_energy}.
\ous{6}{First, \ous{8}{as \gls{nrh} decreases,} the DRAM energy overhead of all studied mitigation mechanisms increases.
Second, \ous{8}{\ous{12}{when} \gls{nrh} decreases from 1K to \ous{12}{20},} \gls{prfm}'s average DRAM energy overhead increases from \ous{12}{\param{3.2}}\% to \ous{12}{\param{4x}}.
Third, \ous{8}{\ous{12}{when} \gls{nrh} decreases from 1K to \ous{12}{20},} \gls{prac}-4's average DRAM energy overhead \ous{12}{significantly} increases from \ous{12}{\param{18.5}}\% to \ous{12}{\param{13x}}.
In contrast, \gls{prac}-Optimistic's (i.e., \gls{prac} \emph{without} increased DRAM timing parameters) average DRAM energy overhead increases from \ous{12}{\param{0.001}}\% to \ous{12}{\param{7x}}.
Therefore, a significant portion of \gls{prac}'s DRAM energy is \om{8}{likely due to} the increased timing parameters.}
\ous{6}{We attribute \ous{7}{the high} increase in DRAM energy overheads of \gls{prfm} and \gls{prac} as \gls{nrh} decreases to 1) their conservative preventive \ous{7}{refresh} thresholds against the wave attack and 2) benign applications triggering \ous{7}{many} preventive \ous{7}{refreshes}.}
\ous{7}{Fourth, \ous{8}{as \gls{nrh} decreases from 1K to \ous{12}{20},} average DRAM energy overheads of Graphene and Hydra increase from \ous{12}{\param{0.01}}\% and \ous{12}{\param{0.3}}\% to \ous{12}{\param{33.2}}\% and \ous{12}{\param{62.7}}\%, respectively.}

\om{8}{We} \ous{6}{conclude that 1) \gls{prac} and \gls{prfm} \om{8}{already} incur \om{8}{relatively} high DRAM energy overheads for \ous{8}{modern} DRAM chips \om{8}{(i.e., \gls{nrh} $=$ 1K)}, 2) energy overheads of all evaluated mitigations mechanisms significantly increase for future DRAM chips that are more vulnerable to read disturbance, \ous{7}{and 3) Graphene and Hydra outperform all \gls{prac} and \gls{prfm} \ous{8}{implementations} at all evaluated \gls{nrh} values across all workloads (except \ous{7}{\gls{prac}-Optimistic at \gls{nrh} values higher than \ous{12}{32}})}.}

\subsection{\ous{7}{Storage Evaluation}}
\label{subsec:storageeval}
\begin{figure}[h]
\centering
\includegraphics[width=\linewidth]{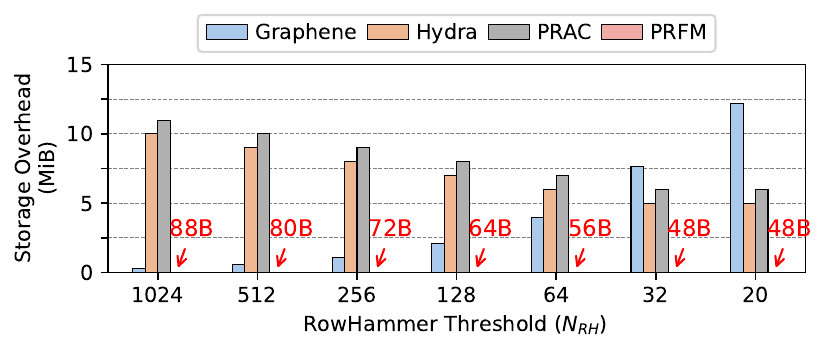}
\caption{Storage used by evaluated read disturbance mitigation mechanisms \om{8}{as a function of RowHammer threshold}}
\label{fig:benign_storage}
\end{figure}

\ous{7}{\figref{fig:benign_storage} presents the storage requirements of the evaluated \agy{2}{read disturbance} mitigation mechanisms as \gls{nrh} decreases.}
\ous{8}{Axes respectively show the \gls{nrh} values (x axis) and storage (y axis) in mebibytes (MiB).}

\ous{7}{From \figref{fig:benign_storage}, we make \ous{8}{four} observations.
First, \ous{8}{as \gls{nrh} decreases from 1K to \ous{12}{20},} Graphene's storage overhead \om{8}{in CPU increases} significantly \om{8}{(by \param{50.3}x)} due to the need to track many more rows.
\ous{8}{Second, as \gls{nrh} decreases from 1K to \ous{12}{20}, \gls{prac} and Hydra's storage \om{9}{overheads} \om{8}{in DRAM} reduce by \ous{12}{\param{45.5}}\% and \ous{12}{\param{50.0}}\%, respectively.}
\ous{8}{We note that while Hydra's cache structure in CPU does \emph{not} change, the overall cache size reduces with \gls{nrh} (by \ous{12}{\param{43.9}}\% from 1K to \ous{12}{20}) as smaller cache entries are sufficient to track activations.}
\ous{8}{Fourth}, \gls{prfm} incurs the least storage overhead \om{8}{in \om{9}{the} CPU} among the evaluated mitigation techniques as it only requires \om{8}{only} one counter per bank.}

\om{8}{We} \ous{7}{conclude that \gls{prac}, \gls{prfm}, and Hydra incur low storage overheads and scale well with decreasing \gls{nrh} values as they either $i$) keep counters in DRAM where a large amount of storage is available at high density or $ii$) require only a small set of counters.}\ouscomment{8}{removed regardless of threshold}
\section{\ous{6}{Performance Degradation Attack}}
\label{sec:evaluation_dos}
\ous{1}{An attacker can take advantage of \gls{prac}
\agy{2}{to mount memory performance (or denial of memory service) attacks\ous{6}{\cite{mutlu2007memory}}}
by triggering many preventive actions \agy{2}{(e.g., back-off signals and \gls{rfm}s)}\gf{5}{.}
\agy{2}{This section presents}
1)~the theoretical maximum downtime of a \agy{2}{\gls{prac}-protected system} and 2)~\agy{2}{simulation results.}}

\ouscomment{1}{This can also be approximated with (time to refresh) / (time to trigger + time to refresh)}
\head{Theoretical Analysis}
\ous{1}{\agy{2}{We calculate} the maximum possible \agy{2}{fraction of time that} preventive actions \agy{2}{take in a \gls{prac}-protected system.}
\agy{2}{First, triggering a back-off signal takes $\aboth\times\trc$, which causes $\bonrefs$ RFM commands, blocking the bank for a time window of $\bonrefs\times\trfm$. Therefore, an attacker can block a DRAM bank for $(\bonrefs\times\trfm)/(\bonrefs\times\trfm+\aboth\times\trc)$ of time.}
We configure \ous{2}{$\aboth$, $\bonrefs$, $\trfm$, and $\trc$ as \ous{12}{\param{1}}, \param{4}, \param{\SI{350}{\nano\second}}, and \param{\SI{52}{\nano\second}}} against an \gls{nrh} of \ous{12}{\param{20}},\agycomment{2}{I think the simplified version of this I put above is correct. Please double-check. Let's put only the numbers that appear in the formula and \gls{nrh}.} \agy{2}{based on \ous{6}{the} \param{DDR5-3200AN} \ous{6}{DRAM timing constraints specified 
in the JEDEC standard}~\cite{jedec2024jesd795c}.}
% 16 GB 3200AN (all ns)
% tRC = 52 (increased due to PRAC)
% tREFW = 32000000
% tREFI = 3900
% tRFC = 295
% tRFM = 350 (increased due to PRAC)
% t_avail = 32000000 - 295*(32000000/3900) = 29579487.1795
% T_attack = 228 * 52 + 4 * 350 = 4144
% t_prevent = 4 * 350 * 29579487.1795 / 4144 = 9993069.99
% t_prevent / t_avail = 9993069.99 / 29579487.1795 = 0.337837838
% By plugging the \param{DDR5-3200AN}~\cite{jedec2024jesd795c} timing preset, 
% we obtain the $t_{available}$, $T_{attack}$, and $t_{prevent}$ as \param{$\approx$29.58ms}, \param{3859ns}, and \param{$\approx$9.04ms}, respectively.
% Based on these values, 
We observe that an attacker can \emph{theoretically} consume \ous{12}{\param{94\%}} of \ous{6}{DRAM} \om{7}{throughput} by triggering back-offs.}

\head{Simulation}
\ous{1}{To understand the \ous{6}{system} \ous{7}{performance} degradation an attacker could \ous{6}{cause} \ous{7}{by hogging the available DRAM throughput with preventive refreshes}, we \agy{2}{simulate} \param{60} \param{four}-core workload mixes of varying memory intensities where \agy{2}{\ous{6}{one} core maliciously}
\agy{2}{hammers} \param{8} rows \ous{2}{in each of} \param{4} banks\agy{2}{.}}\footnote{\ous{6}{We experimentally found these values to yield the highest performance overhead \om{7}{for} \gls{prac} \om{7}{in} our system configuration. We open-source our attacker trace generator with the rest of our implementation to \om{7}{aid} reproducibility~\cite{ramulator2github}.}}

\agy{2}{Our results for \gls{nrh} values of 128, 64, 32, and \ous{12}{20} show that \gls{prac} \ous{6}{reduces system \ous{7}{performance (based on the weighted speedup metric~\cite{eyerman2008systemlevel, snavely2000symbiotic})} on average (maximum)} by \ous{12}{\param{18.4}\% (\param{29.0}\%), \param{22.9}\% (\param{34.0}\%), \param{49.2}\% (\param{75.0}\%), and \param{86.8}\% (\param{94.6}\%)} \ous{8}{with a maximum slowdown\om{9}{\cite{kim2010thread, jattke2022blacksmith}} on a single application of \ous{12}{\param{64.5}\%, \param{66.8}\%, \param{75.0}\%, and \param{97.7}\%}}, respectively.
These results indicate that \agy{6}{memory performance attacks can exploit} \gls{prac} 
and \agy{6}{future research should tackle} \gls{prac}'s performance overheads \om{7}{to avoid such denial of service attacks}.}
\section{Summary and Future Research Directions}
\label{sec:implications}

We show that \gls{prac} ensures \om{6}{secure} operation even for \om{8}{very} low \gls{nrh} values \gf{6}{(e.g.,} \ous{6}{as low as \ous{12}{20}}\om{7}{,} \gf{6}{see \secref{sec:configurationandsecurity})}.\ouscomment{6}{Revised our claims}
However, \gls{prac} still incurs \om{6}{high} performance and energy overheads \ous{6}{especially at low \gls{nrh} values (e.g., \ous{12}{$\leq$ 32}, \om{7}{see \secref{subsec:perfeval}})}, which can be maliciously exacerbated to mount memory performance attacks \om{7}{(\secref{sec:evaluation_dos})}.
Therefore, reducing \gls{prac}'s performance \om{6}{and energy overheads} \om{7}{and avoiding its denial of service vulnerability are} still important research \om{7}{problems}.

We identify \om{7}{at least} four directions to explore.
\ous{6}{A} first direction is to reduce the \gls{trp} and \gls{trc} timing constraints that increase when \gls{prac} is enabled.
\ous{6}{These increased DRAM timing parameters incur non-negligible system performance overheads even at high \gls{nrh} values.}
This reduction can be done by 1)~leveraging large safety margins \om{8}{associated with timing parameters} \ous{6}{(as shown in~\cite{chandrasekar2014exploiting, lee2015adaptive, chang2016understanding, chang2017understanding, chang2017understandingphd, kim2018solar, yaglikci2022understanding, wang2018reducing, lee2017design, olgun2021quac, kim2019drange, yaglikci2022hira, kim2018dram, gao2019computedram, gao2022frac, orosa2021codic, talukder2019prelatpuf})} or 2)~modifying the DRAM circuitry to separate the counters from data arrays to parallelize \ous{6}{row activation counter} accesses~\cite{bennett2021panopticon}.
\ous{6}{A} second direction is to overlap the latencies of preventive refreshes and other memory operations.
\ous{6}{\om{8}{A} workload triggers \om{7}{more} preventive actions as \gls{nrh} decreases, as even \om{8}{a} benign application \om{8}{starts} activating DRAM rows \ous{7}{too many (i.e., closer to or more than \gls{nrh})} \ous{8}{times}.}\ouscomment{8}{moved the parentheses behind, deleted the ``times'' in parantheses}
\ous{6}{Overlapping the latencies of preventive refreshes} is possible by 1)~leveraging subarray-level parallelism~\cite{kim2012acase, chang2014improving, yaglikci2022hira} or 2)~\om{7}{eliminating} \ous{6}{the blocking nature of preventive refreshes}~\cite{nguyen2018nonblocking, hassan2022acase}.
\ous{6}{A} third direction is to leverage the significant variation in read disturbance vulnerability across DRAM rows to avoid overprotecting the vast majority of the rows\ous{6}{\cite{yaglikci2024spatial,orosa2021deeper}}.
This direction requires profiling a given chip with fast, accurate, and comprehensive \om{7}{(and likely online\om{8}{\cite{patel2017reaper,qureshi2015avatar,liu2013experimental,khan2014efficacy}})} profiling methodologies, which addresses several aspects, including RowHammer's complex interaction with temperature~\cite{orosa2021deeper, orosa2022spyhammer} and new read disturbance phenomena like RowPress~\cite{luo2023rowpress}.
\ous{2}{\ous{6}{A} fourth direction is to defend against malicious attackers that exploit preventive refreshes.}
\ous{6}{Attackers can trigger increasing amounts of preventive refreshes as \gls{nrh} decreases, allowing a new attack vector to conduct memory performance attacks\om{7}{\cite{mutlu2007memory}}.}
\ous{2}{\ous{6}{Preventing these performance attacks} \ous{7}{may be} possible by accurately detecting and throttling \ous{6}{workloads that trigger many preventive refreshes}~\cite{yaglikci2021blockhammer, canpolat2024leveraging}.}\omcomment{8}{Last sentence is too optimistic. It does not consider benign workloads}
\section{Related Work}
\label{sec:relatedwork}

This is the first work that rigorously analyses the security and performance of \gls{prac}, a key feature introduced in the \gf{5}{latest JEDEC} \ous{6}{DDR5} DRAM specification~\cite{jedec2024jesd795c}.
\copied{BreakHammer}{\ous{6}{\secref{sec:evaluation} qualitatively and quantitatively compares \gls{prac} to several \om{7}{prominent} RowHammer mitigation mechanisms~\cite{park2020graphene,qureshi2022hydra,kim2014flipping}.
\atb{6}{There \om{7}{are various other mitigation} mechanisms that can be implemented in the memory controller~\mcBasedRowHammerMitigations{} or in the DRAM chip~\inDRAMRowHammerMitigations{}. We leave \om{7}{a} rigorous \om{8}{comparison} of \gls{prac} to this broader set of RowHammer mitigation \om{7}{techniques to} future work.}}}

\section{Conclusion}
\label{sec:conclusion}

\omcomment{5}{After reading all this, I still do not get a clear conclusion of where the perf degradation is coming from. Is it the increased timing parameters? Is it the back off signal latency or frequency? What is it? Give insight. That is the most important function of a paper and its results.}

\ous{6}{We presented the first rigorous security, performance, energy, and cost analyses of \ous{7}{Per Row Activation Counting (\gls{prac})}, \om{7}{the state-of-the-art RowHammer mitigation technique that is recently adopted by industry in the DDR5 standard~\cite{jedec2024jesd795c}}.
We show that \gls{prac}\ouscomment{7}{changed conclusion order}
1)~has non-negligible overheads due to increased DRAM timing parameters for today's DRAM chips,
2)~incurs significant system performance and DRAM energy overheads by triggering increasingly \om{7}{more} back-off requests for future DRAM chips with \om{7}{higher} read disturbance vulnerabilities,
\ous{7}{3)~can be used \om{8}{as a} memory performance attack \om{8}{vector} to consume a significant portion of the available DRAM throughput \om{8}{and thus degrade overall system performance}, and}
4)~provides secure operation for \gls{nrh} values as low as \ous{12}{\param{20}}.}

\om{7}{We conclude that more research is needed to} \ous{7}{improve \gls{prac} by
$i$) reducing the high system performance and DRAM energy overheads due to increased DRAM timing parameters,
$ii$) solving the exacerbated performance \om{8}{impact} as \gls{nrh} decreases, and
$iii$) stopping its preventive refreshes from being exploited by memory performance attacks.}

% \agy{1}{In this work, we present the first rigorous security, performance, energy, and cost analyses of
% the% state-of-the-art 
% on-DRAM-die read disturbance mitigation method, widely known as \gls{prac}\gf{5}{,} with respect to its description in the updated \gf{5}{JEDEC} DDR5 specifications.}
% % and performance analyses of \agy{1}{a key feature of state-of-the-art DRAM standards:} \gls{prac}\agy{1}{, which provides the DRAM chip with the necessary time window to perform preventive refreshes when necessary. \gls{prac} is important for practical and low-overhead read disturbance mitigation for modern and future DRAM chips.}
% % To do this analysis, we 
% % 1)~formally define an adversarial access pattern which represents the worst-case for \gls{prac}, 
% % 2)~investigate \gls{prac}'s different configurations and their security guarantees, and 
% % 3)~evaluate \gls{prac}'s performance, energy, and cost overheads.
% We show that \gls{prac}
% 1)~can provide robust operation for a hammer threshold as low as \param{10} activations per aggressor row and 
% 2)~incurs less than \param{22.3\%} and an average of \param{17.4\%} performance overhead across \param{60} randomly chosen benign workload mixes for modern and future DRAM chips.

\section*{\om{10}{Acknowledgments}} {
\ous{6}{We thank the anonymous reviewers of DRAMSec \om{7}{2024} for \om{7}{the positive} feedback.
We thank the SAFARI Research Group members for valuable feedback and the stimulating scientific and intellectual environment they provide.
We acknowledge the generous gift funding provided by our industrial partners (especially Google, Huawei, Intel, Microsoft, VMware), which has been instrumental in enabling the research we have been conducting on read disturbance in DRAM \om{7}{since 2011~\cite{mutlu2023retrospective}}.
This work was \om{7}{also} in part supported by the Google Security and Privacy Research Award and the Microsoft Swiss Joint Research Center.}
}

\balance
\bibliographystyle{IEEEtran}
\bibliography{refs}

% \pagebreak
% \onecolumn
% \appendices 
% \input{sections/a_latencyevaluation}
% \input{sections/b_performanceattack}

\end{document}